\documentclass[preprint,12pt]{elsarticle}

\usepackage{amssymb}

\def\be{\begin{equation}}
\def\ee{\end{equation}}
\def\bea{\begin{array}}
\def\eea{\end{array}}
\def\beqa{\begin{eqnarray}}
\def\eeqa{\end{eqnarray}}
\def\beqas{\begin{eqnarray*}}
\def\eeqas{\end{eqnarray*}}

\def\bp{\begin{picture}}
\def\ep{\end{picture}}
\def\bc{\begin{center}}
\def\ec{\end{center}}
\def\bfig{\begin{figure}}
\def\efig{\end{figure}}

\def\bit{\begin{itemize}}
\def\eit{\end{itemize}}
\def\nn{\nonumber}
\def\f{\frac}

\def\[{\left[}
\def\]{\right]}
\def\({\left(}
\def\){\right)}

\def\..{\left.}
\def\.{\right.}
\def\tl{\tilde}
\def\ra{\rightarrow}

\def\tm{\times}

\def\al{\alpha}

\def\ep{\epsilon}

\def\de{\delta}

\def\pr{\prime}

\journal{Nuclear Physics B}

\begin{document}

\begin{frontmatter}



\title{Supersymmetry Breaking Scalar Masses and Trilinear Soft Terms From High-Dimensional Operators in $E_6$ SUSY GUT}


\author{Fei Wang$^{1,2}$}

\address{1. Department of Physics, Zhengzhou University, Zheng Zhou, 450001, China \\
2. School of Physics, Monash University, Melbourne Victoria 3800, Australia}

\begin{abstract}
In the GmSUGRA scenario with the
higher-dimensional operators containing the GUT Higgs fields,
we systematically studied the supersymmetry breaking scalar masses,
SM fermion Yukawa coupling terms,
and trilinear soft terms in the $E_6$ SUSY GUT model where
the gauge symmetry is broken down to the
$SO(10)\tm U(1)$ gauge symmetry,
$SU(3)_C\times SU(3)_L \times SU(3)_R$
gauge symmetry, $SU(6)\times SU(2)_a (a={\rm L,R,X})$
gauge symmetry, flipped $SU(5)$ gauge symmetry etc. In addition, we considered the scalar and
gaugino mass relations, which can be preserved from
the GUT scale to the electroweak scale
under one-loop RGE running, in the $SU(3)_C\times SU(3)_L \times SU(3)_R$
model arising from the $E_6$ model. With such relations, we may distinguish the
mSUGRA and GmSUGRA scenarios if we can measure the supersymmetric
particle spectrum at the LHC and ILC.
\end{abstract}


\end{frontmatter}


\section{Introduction}

 Supersymmetry naturally solves the
gauge hierarchy problem of the Standard Model (SM). The unification
of the three gauge couplings $SU(3)_C, SU(2)_L$ and $U(1)_Y$ in the
supersymmetric Standard Model at about $2\tm 10^{16}$
GeV~\cite{Ellis:1990zq} strongly suggests the existence of Grand
Unified Theories (GUTs). In addition,
GUT models such as $SU(5)$~\cite{Georgi:1974sy},
$SO(10)$~\cite{so10}, and superstring-inspired $E_6$~\cite{e61,e62} models etc \cite{fei0} give us deep insights into the
other SM problems such as the emergence of the fundamental forces,
the assignments and quantization of their charges,
the fermion masses and mixings, and beyond.
Although supersymmetric GUTs are
attractive it is challenging to test them
at the Large Hadron Collider (LHC), the future
International Linear Collider (ILC), and other experiments.

In traditional supersymmetric SMs,
supersymmetry is broken in the hidden sector and
the supersymmetry breaking effects can be mediated
to the observable sector via gravity~\cite{mSUGRA},
gauge interactions~\cite{Ellis:1984bm, gaugemediation},
or super-Weyl
anomaly~\cite{anomalymediation, UVI-AMSB, D-AMSB}, or other mechanisms.
Recently, considering GUTs with higher-dimensional
operators~\cite{Ellis:1984bm, Ellis:1985jn, Hill:1983xh,
Shafi:1983gz, Drees:1985bx,
Anderson:1999uia, Chamoun:2001in, Chakrabortty:2008zk, Martin:2009ad,
Bhattacharya:2009wv, Feldman:2009zc, Chamoun:2009nd,stefan}
and F-theory GUTs with $U(1)$ fluxes~\cite{Vafa:1996xn,
Donagi:2008ca, Beasley:2008dc, Donagi:2008kj,
Font:2008id, Jiang:2009zza, Blumenhagen:2008aw, Jiang:2009za,
Li:2009cy, Leontaris:2009wi, Li:2010mr},
generalized mSUGRA (GmSUGRA) scenario is proposed \cite{Li:2010xr} in which the gaugino mass relations are studied
and their indices are defined. In our previous works \cite{fei1,fei2}, we discuss (in the context of GmSUGRA) the supersymmetry breaking
scalar masses and trilinear soft terms in SU(5) and SO(10) GUT models
with various higher dimensional Higgs fields. It is also interesting to discuss in $E_6$ SUSY GUT model the supersymmetry breaking
scalar masses and trilinear soft terms from non-renormalizable Kahler potential and non-renormalizable superpotential.

 The exceptional group $E_6$ has been proposed as an attractive unification group with several desirable features: 1) $E_6$ was the
next natural anomaly-free choice for a GUT group after $SO(10)$; 2) all the basic fermions of one generation belong to a single irreducible representation
{\bf 27}. We know that within the context of heterotic superstring theory in ten dimensions, gauge and gravitational anomaly cancelation was found to occur only
for the gauge groups $SO(32)$ or $E_8 \tm E_8$ \cite{GS}.  Compactification on a Calabi-Yau manifold with an SU(3) holonomy results in the breaking
$E_8\ra SU(3) \tm  E_6$ with the $SU(3)$ gauge field becoming the spin connection on the compactified space. This result inspired the current interests in $E_6$ GUT \cite{e6cite1,e6cite2,e6cite3,e6cite4,e6cite5,e6cite6,e6cite7}.

In this paper, we consider the supersymmetry breaking
scalar masses and trilinear soft terms from non-renormalizable Kahler potential and non-renormalizable superpotential in $E_6$ SUSY GUT.
We systematically calculate the supersymmetry breaking scalar masses,
SM fermion Yukawa coupling terms,
and trilinear soft terms in $E_6$ models where
the gauge symmetry is broken down to the $SO(10)\tm U(1)$ gauge symmetry, flipped SO(10)\cite{fso101,fso102,fso103,fso104} gauge symmetry,
$SU(3)_C\times SU(3)_L \times SU(3)_R$ gauge symmetry, $SU(6)\tm SU(2)_X$ \cite{su61,su62,su63} gauge symmetry, flipped $SU(5)\times U(1)_X$ gauge
symmetry~\cite{smbarr, dimitri, AEHN-0}. We should note that we investigates in this work only the group-theoretical
necessities for such a breaking, but no dynamical model is constructed to give the symmetry-breaking
vacuum expectation value (VEV). Besides, in our work we consider basically one single spontaneously symmetry breaking step for $E_6$. As a result, no investigations about the running of the coupling in general and possible constraints from perturbativity have been made in this work.
We examine the scalar and
gaugino mass relations, which are valid from
the GUT scale to the electroweak scale
under one-loop renormalization group running in the $SU(3)_C\times SU(3)_L \times SU(3)_R$
models arising from the $E_6$ GUT model.
With these relations, we may distinguish the
mSUGRA and GmSUGRA scenarios if the supersymmetric
particle spectrum can be measured at the LHC and ILC.

   This paper is organized as follows. In Section~\ref{sec-0},
we briefly review four-dimensional $E_6$ GUTs and its symmetry breaking chains.
In Section~\ref{sec-1}, we explain the general gravity mediated
supersymmetry breaking. We derive the scalar
masses in Section~\ref{sec-2}, and the SM fermion Yukawa
coupling terms and trilinear soft terms in Section~\ref{sec-3}.
In Section~\ref{sec-4} we consider
the scalar and gaugino mass relations.
Section~\ref{sec-5} contains our conclusions.
\section{ Brief Review of Grand Unified Theories}
\label{sec-0}

In this Section we explain our conventions.
In supersymmetric SMs,
we denote the left-handed quark doublets, right-handed
up-type quarks, right-handed down-type quarks,
left-handed lepton doublets, right-handed neutrinos
and right-handed charged leptons as $Q_L^i$, $(U^c_L)^i$, $(D^c_L)^i$,
$L_L^i$, $(N^c_L)^i$, and $(E^c_L)^i$, respectively. Also, we denote
one pair of Higgs doublets as $h_u$ and $h_d$, which give masses
to the up-type quarks/neutrinos and the down-type quarks/charged
leptons, respectively.

First, we briefly review the $E_6$ GUT model. $E_6$ can break into gauge group $SU(3)_C\tm SU(3)_L\tm SU(3)_R$ , $SU(6)\tm SU(2)_a$ (a=L,R,X)
gauge symmetry, $SO(10)$ gauge symmetry, flipped SO(10) gauge symmetry, flipped SU(5) gauge symmetry, Pati-Salam $SU(4)_c\tm SU(2)_L\tm SU(2)_R$ gauge symmetry,
$SU(3)_C\tm SU(2)_L\tm SU(2)_R\tm U(1)_1 \tm U(1)_2$ gauge symmetry.
Each generation of standard model matter contents are filled into ${\bf 27}$ dimensional representations of $E_6$ GUT group.
Depending on different gauge symmetry breaking chains, the standard model matter contents are filled differently.
\begin{itemize}
\item $SU(3)_C\tm SU(3)_L\tm SU(3)_R$

Under $SU(3)_C\tm SU(3)_L\tm SU(3)_R$ gauge symmetry, the ${\bf 27}$ and ${\bf 78}$ dimensional representation of $E_6$ are decomposed \cite{Slansky:1981yr}
\beqa
{\bf 27}&=&{\bf (~3,~{3},~1)}\oplus {\bf (~\bar{3},~1,~\bar{3})}\oplus {\bf (~1,~\bar{3},~3)}~,\\
{\bf 78}&=&{\bf (~8,~1,~1)}\oplus {\bf (~1, ~8,~1)}\oplus {\bf (~1,~1,~8)}\nn\\
&&\oplus {\bf (~3,~\bar{3},~\bar{3})}\oplus {\bf (\bar{3},~3,~{3})}.
\eeqa
 The filling of the standard model matter contents in terms of gauge group $SU(3)_C\tm SU(3)_L\tm SU(3)_R$
 \beqa
 \small
X_L^a{\bf (~3,~{3},~1)} &\sim& \left(\bea{c}u_L\\d_L\\D_L\eea\right),~~~
(X_L^c)^a{\bf (~\bar{3},~1,~\bar{3})} \sim \left(\bea{c}u_L^c\\d_L^c\\D_L^c\eea\right),~\nn\\
N^a{\bf (~1,~\bar{3},~3)}&\sim& \left(\bea{ccc}H_1^0&H_2^+&e_L^c\\H_1^-&H_2^0&-\nu_L^c\\e_L&-\nu_L&n_0\eea\right),
 \eeqa\normalsize
where $(a=1,2,3)$ for three families. The breaking of gauge group $E_6$ into $SU(3)_C\tm SU(3)_L\tm SU(3)_R$ is achieved by ${\bf 650}$ dimensional Higgs field.\footnote{
The decomposition of ${\bf 650}$ dimensional Higgs into $SU(3)_C\tm SU(3)_L\tm SU(3)_R$ quantum numbers is
\beqa\small
{\bf 650}&=&{\bf (~1,~1,~1)}\oplus {\bf (~1, ~1,~1)}\oplus {\bf (~8,~1,~1)}\oplus {\bf (~1,~8,~1)}\oplus {\bf (~1,~1,~8)}
\oplus {\bf (~\bar{3},~{3},~{3})}\nn\\&\oplus& {\bf (\bar{3},~3,~{3})}\oplus {\bf (~{3},~\bar{3},~\bar{3})}\oplus {\bf (~{3},~\bar{3},~\bar{3})}
\oplus {\bf (~{3},~{6},~\bar{3})}\oplus {\bf (~{3},~\bar{3},~{6})}\oplus {\bf (~\bar{3},~\bar{6},~{3})}\nn\\
&\oplus& {\bf (~\bar{3},~{3},~\bar{6})}\oplus {\bf (~\bar{6},~\bar{3},~\bar{3})}\oplus {\bf (~{6},~{3},~{3})}\oplus {\bf (~8,~8,~1)}
\oplus {\bf (~1,~8,~8)}\nn\\&\oplus& {\bf (~8,~1,~8)}~.
\eeqa\normalsize}
  To break gauge symmetry $SU(3)_C\tm SU(3)_L\tm SU(3)_R$ into its subgroup $SU(3)_C\tm SU(2)_L\tm SU(2)_R \tm U(1)_1\tm U(1)_2$,
we can use ${\bf 650}$ dimensional Higgs fields to acquire ${\bf (~1,~8,~8)}$ term VEVs. It is also possible to use ${\bf 27,\overline{27}}$
dimensional representation Higgs to achieve the second stage symmetry breaking into the left-right gauge group $SU(3)_C\tm SU(2)_L\tm SU(2)_R \tm U(1)_{B-L}$.

\item $SO(10)\tm U(1)$

The fundamental representation ${\bf 27}$ and adjoint representation ${\bf 78}$ of $E_6$ can be decomposed in term of $SO(10) \tm U(1)$
\beqa
{\bf 27}&=&{\bf 16}_1\oplus {\bf 10}_{-2}\oplus {\bf 1}_{4}~,\\
{\bf 78}&=&{\bf 45}_0\oplus {\bf 16}_{-3}\oplus {\bf \overline{16}}_{3}\oplus {\bf 1}_{0}~.
\eeqa

 The filling of standard model contents is different between the flipped $SO(10)$ scenario and the $U(1)$ extension of ordinary $SO(10)$ scenario.
\begin{itemize}
\item $U(1)$ extension of ordinary $SO(10)$

   In this scenario, the standard model matter contents are filled in ${\bf 16}$ dimensional spinor representation (decomposed in Georgi-Glashow $SU(5)\tm U(1)$)
 \beqa
 {\bf 16_1}&=&{\bf (10_{\rm Q_L,U_L^c,E_L^c},~\bar{5}_{\rm D_L^c,L_L},~1_{N_L^c})}~,\nn\\
 {\bf 10_{-2}}&=&{\bf ( 5_H ,\bar{5}_H )}~,\nn\\
 {\bf 1_{-4}}&=&{\bf 1_S}~.
 \eeqa

Flipped $SU(5)\times U(1)_{X}$ model~\cite{smbarr, dimitri, AEHN-0} can also be embedded into $SO(10)$.

   In this scenario, the standard model matter contents are filled as
 \beqa
 {\bf 16_1}&=&{\bf (10_{\rm Q_L,D_L^c,N_L^c},~\bar{5}_{\rm U_L^c,L_L},~1_{E_L^c})}~\nn\\
 {\bf 10_{-2}}&=&{\bf ( 5_V ,\bar{5}_V )}~,\nn\\
 {\bf 1_{-4}}&=&{\bf 1_S}~.
 \eeqa

To break the flipped $SU(5)$ GUT and electroweak gauge symmetries, we introduce two pairs
of Higgs fields whose quantum numbers under $SU(5)\times U(1)_X$ are
\begin{eqnarray}
H={\mathbf{(10, 1)}}~,~~{\overline{H}}={\mathbf{({\overline{10}},
-1)}}~,~~
h={\mathbf{(5, -2)}}~,~~{\overline h}={\mathbf{({\bar {5}}, 2)}}~,~\,
\label{Higgse1}
\end{eqnarray}
where $h$ and ${\overline h}$ contain the Higgs doublets
$h_d$ and $h_u$, respectively. In flipped SU(5), Doublet-Triplet splitting problems can be solved via the elegant missing partner mechanism.
This mechanism is however spoiled if we embed flipped SU(5) into SO(10) \cite{f5so101,f5so102,fei3}.
\item Flipped $SO(10)$

   Flipped SO(10) is introduced in \cite{fso101,fso102,fso103,fso104} to keep the elegant missing partner mechanism when embedding
flipped SU(5) into $SO(10)\tm U(1)$. In this scenario, the standard model matter contents (with extra exotic particles) are filled as
 \beqa
 {\bf 16_1}&=&{\bf (10_{\rm Q_L,D_L^c,N_L^c},~\bar{5}_V,~1_S)}~,\\
 {\bf 10_{-2}}&=&{\bf ( 5_V ,\bar{5}_{\rm U_L^c,L_L} )}~,\\
 {\bf 1_{-4}}&=&{\bf 1_{\rm E_L^c}}~,
 \eeqa
in which we flip ${\bf \bar{5}_{\rm U_L^c,L_L}}$ with ${\bf \bar{5}_V}$, ${\bf 1_{\rm E_L^c}}$ with ${\bf 1_S}$ with respect to
ordinary embedding of flipped $SU(5)$ into $SO(10)$.
\end{itemize}

\item $SU(6)\tm SU(2)_a$(a=L,R,X)

  In the simplest grand unifying group SU(5), natural implementation of doublet triplet splitting seems to require the
use of the relatively large representations ${\bf 50},{\bf \overline{50}}$ and ${\bf 75}$. The search for a simpler solution has lead various authors
to consider the extension of the gauge symmetry to SU(6) \cite{su61,su62,su63} which allows for more possibilities:
(i)The light Higgs doublets emerge as the pseudo-Goldstone bosons of a broken global symmetry of the superpotential \cite{gift1,gift2}.
(ii) the sliding singlet mechanism where the desired VEV pattern follows automatically from the conditions of the supersymmetric minima condition.
In the context of SU(5) model, there are severe difficulties due to radiative corrections which actually lift the MSSM doublet masses to an intermediate scale.
If instead one considers an embedding of the SU(5) model into the SU(6) group, then the problems associated with radiative instability of doublet-triplet
splitting can be cured \cite{missvev1,missvev2}.

The fundamental representation ${\bf 27}$ and adjoint representation ${\bf 78}$ of $E_6$ can be decomposed in term of $SU(6) \tm SU(2)$
\beqa
{\bf 27}&=&{\bf (~\bar{6},~{2})}\oplus {\bf (~15,~1)}~,\\
{\bf 78}&=&{\bf (~35,~1)}\oplus {\bf (~1, ~3)}\oplus {\bf (~20,~2)}~.
\eeqa
  The breaking of $E_6$ into $SU(6)\tm SU(2)_a$(a=L,R,X) can be achieved by ${\bf 650}$ dimensional Higgs field
whose decomposition reads
\beqa
\small
{\bf 650}&=&{\bf (~1,~1)}\oplus {\bf (~1, ~35)}\oplus {\bf (~2,~20)}\oplus {\bf (~3,~35)}\oplus {\bf (~2,~70)}\nn\\&&
\oplus{\bf (~2,~\overline{70})}\oplus {\bf (~1, ~189)}.
\eeqa
Different choice of the $SU(2)_a$ leads to different filling of the standard model matter contents
\begin{itemize}
\item $E_6\ra SU(6)\tm SU(2)_X\ra SU(5)\tm U(1)\tm SU(2)_X$:

The decomposition of ${\bf 27}$ representation in terms of $SU(5)\tm SU(2)_X$
\beqa
{\bf (~15,~1)}&=&{\bf (~10,~1)\oplus(~5,~1)}~,\\
{\bf (~~\bar{6},~2)}&=&{\bf (~\bar{5},~2)\oplus(~1,~2)}~.
\eeqa
We identify the matter contents
\beqa
{\bf (~10,~1)}&\supset&(~U_L,~U_L^c,~D_L, E_L^c)~,\\
{\bf (~\bar{5},~2)}&\supset&(~D_L^c, ~E_L,~N_L)~,\\
{\bf (~1,~2)}&\supset& N_L^c~.
\eeqa

\item $E_6\ra SU(6)\tm SU(2)_L \ra SU(4)_c\tm SU(2)_L\tm SU(2)_R \tm U(1)_2$:

This symmetry breaking chain was proposed in \cite{su6ps}.
The decomposition of ${\bf 27}$ representation of $E_6$ in terms of gauge group $SU(4)_c\tm SU(2)_R\tm SU(2)_L$ reads
\beqa
{\bf (~15,~1)}&=&{\bf (~6,~1,~1)\oplus(~1,~1,~1)\oplus(~4,~2,~1)}~,\\
{\bf (~~\bar{6},~2)}&=&{\bf (~\bar{4},~1,~2)\oplus(~1,~\bar{2},~2)}~.
\eeqa

We identify the matter contents
\beqa
{\bf (~4,~2,~1)}&\supset&(~U_R,~D_R,~E_R,~N_R)~,\\
{\bf (~\bar{4},~1,~2)}&\supset&(~U_R^c,~D_R^c,~E_R^c,~N_R^c)~.
\eeqa

\item $E_6\ra SU(6)\tm SU(2)_R \ra SU(4)_c\tm SU(2)_L\tm SU(2)_R \tm U(1)_2$:

The decomposition of ${\bf 27}$ representation in terms of gauge group $SU(4)_c\tm SU(2)_L\tm SU(2)_R$ reads
\beqa
{\bf (~15,~1)}&=&{\bf (~6,~1,~1)\oplus(~1,~1,~1)\oplus(~4,~2,~1)}~,\\
{\bf (~~\bar{6},~2)}&=&{\bf (~\bar{4},~1,~2)\oplus(~1,~\bar{2},~2)}~.
\eeqa

We identify the matter contents
\beqa
{\bf (~4,~2,~1)}&\supset&(~U_L,~D_L,~e_L, \nu_L)~,\\
{\bf (~\bar{4},~1,~2)}&\supset&(~U_L^c,~D_L^c,~e_L^c, \nu_L^c)~.
\eeqa
\end{itemize}

\end{itemize}

\section{General Gravity Mediated Supersymmetry Breaking}
\label{sec-1}
The supegravity scalar potential can be written as~\cite{mSUGRA}
\beqa {V}=M_*^4e^G\[G^i(G^{-1})^j_iG_j-3\]+\f{1}{2} {\rm
Re}\[(f^{-1})_{ab}\hat{D}^a\hat{D}^b\] ~,~\eeqa
where $M_*$ is the fundamental scale,
D-terms are
\beqa
\hat{D}^a{\equiv}-G^i(T^a)_i^j\phi_j=-\phi^{j*}(T^a)_j^iG_i~,\eeqa
and the K\"ahler function $G$ as well as its
derivatives and metric $G_i^j$ are
\beqa
G&{\equiv}&\f{K}{M_{*}^2}+\ln\(\f{W}{M_{*}^3}\)+\ln\(\f{W^*}{M_*^3}\)~,\\
G^i&=&\f{\delta G}{\de \phi_i}~,~~G_i=\f{\de G}{\de
\phi_i^*}~,~~G_i^j=\f{\de^2 G}{\de\phi^*_i\de\phi_j}~,~\,
\eeqa
where $K$ is K\"ahler potential and $W$ is superpotential.

Because the gaugino masses have been studied
previously~\cite{Li:2010xr}, we only
consider the supersymmetry breaking scalar masses and trilinear
soft terms in this paper.
To break supersymmetry, we introduce a chiral superfield $S$ in
the hidden sector whose $F$ term acquires a vacuum expectation
value (VEV), ${\it i.e}$, $\langle S \rangle = \theta^2 F_S$.
To calculate the scalar masses and trilinear
soft terms, we consider the following superpotential
and K\"ahler potential
\begin{eqnarray}
W &=& {1\over 6} y^{ijk} \phi_i \phi_j \phi_k +
\alpha {S \over {M_*}}
\left( {1\over 6} y^{ijk} \phi_i \phi_j \phi_k \right)~,~\,
\end{eqnarray}
\begin{eqnarray}
K &=& \phi_i^{\dagger} \phi_i + \beta {{S^{\dagger} S}\over {M^2_*}}
\phi_i^{\dagger} \phi_i
~,~\,
\end{eqnarray}
where $y^{ijk}$, $\alpha$, and $\beta$ are Yukawa couplings.
Thus, we obtain the universal supersymmetry breaking
scalar mass $m_0$ and
trilinear soft term $A$ of mSUGRA
\begin{eqnarray}
m^2_0~=~ \beta {{|F_S|^2}\over {M^2_*}}~,~~~ A_{ijk}~=A_0~y_{ijk}=\(\alpha {{F_S}\over
{M_*}}\)y_{ijk}~.~\,
\end{eqnarray}

When we break the GUT gauge symmetry by giving VEV to the
Higgs field $\Phi$, we can have the general superpotential
and K\"ahler potential
\begin{eqnarray}
W &=& {1\over 6} y^{ijk} \phi_i \phi_j \phi_k
+ {1\over 6} \left( h^{ijk} {{\Phi}\over M_*} \phi_i \phi_j \phi_k
\right)
+ \alpha {S \over {M_*}}
\left( {1\over 6} y^{ijk} \phi_i \phi_j \phi_k \right) \nonumber \\
&& + \alpha' {T \over {M_*}}
\left( {1\over 6} y^{ijk} {{\Phi}\over M_*} \phi_i \phi_j \phi_k \right)
~,~\,
\end{eqnarray}
\begin{eqnarray}
K &=& \phi_i^{\dagger} \phi_i + {1\over 2} h' \phi_i^{\dagger}
\left({{\Phi}\over M_*}+
{{{\Phi}^{\dagger}} \over M_*} \right) \phi_i
+\beta {{S^{\dagger} S}\over {M^2_*}} \phi_i^{\dagger} \phi_i
\nn\\&&+
\f{1}{2}{\beta'}^{\Phi} {{T^{\dagger} T}\over {M^2_*}} \phi_i^{\dagger}
\left({{\Phi}\over M_*}+
{{{\Phi}^{\dagger}} \over M_*} \right) \phi_i
~,~\,
\end{eqnarray}
where $h^{ijk}$, $\alpha'$, ${\beta'}^{\Phi}$ and $h'$ are Yukawa couplings, and
$T$ can be $S$ or another chiral superfield with non-zero $F$ term,
${\it i.e}$, $\langle T \rangle = \theta^2 F_T$.
Therefore, after the GUT gauge symmetry is broken
by the VEV of $\Phi$, we obtain the
non-universal supersymmetry breaking scalar masses and trilinear soft
terms, which will be studied in the following.
For simplicity, we assume $h'=0$ in the following discussions
since we can redefine the fields and the SM fermion Yukawa couplings.

\section{Non-Universal Soft masses for sfermions in $E_6$ SUSY GUT}
\label{sec-2}
 We know that the matter contents in $E_6$ GUT are fitted into ${\bf 27}$ dimensional representations. Thus the non-minimal kinetic terms for matter contents in $E_6$ GUT requires the group tensor production decomposition \cite{Slansky:1981yr}
\beqa {\bf \overline{27}}\otimes{\bf 27}&=&{\bf 1}\oplus {\bf
78}\oplus{\bf 650}~. \eeqa
So in order to construct general gauge invariant non-renormalizable Kahler potential terms, we need to consider Higgs in ${\bf 78}$ and ${\bf 650}$ dimensional representations.

\subsection{$E_6$ To $SO(10)\tm U(1)_1$ Model}
The fundamental representation ${\bf 27}$ and adjoint representation ${\bf 78}$ of $E_6$ can be decomposed in term of $SO(10) \tm U(1)$
\beqa
{\bf 27}&=&{\bf 16}_1\oplus {\bf 10}_{-2}\oplus {\bf 1}_{4}~,\\
{\bf 78}&=&{\bf 45}_0\oplus {\bf 16}_{-3}\oplus {\bf \overline{16}}_{3}\oplus {\bf 1}_{0}~.
\eeqa
 The ${\bf 78}$ dimensional representation Higgs can acquire Vacuum Expectation Values (VEVs)
which break $E_6$ into $SO(10)\tm U(1)$. Such VEVs can be written as $27\tm 27$ matrix as follows
\beqa
<\Phi>^{\bf 78}=\f{\hat{v}_{\bf 78}}{2\sqrt{6}}{\rm diag}(\underbrace{~1,\cdots,~1}_{16},\underbrace{-2,\cdots,-2}_{10},~4)~.
\eeqa
with normalization factor
\beqa
c=Tr(<\Phi>^2)=T({\bf 27})=3~.\nn
\eeqa
We normalize the VEVs with $Tr(T^aT^b)=T(r)\delta^{ab}$, so that same results will be obtained when the same VEVs are written as different $n\times n$ matrix forms.

The ${\bf 650}$ dimensional Higgs can also acquire Vacuum Expectation Values (VEVs)
which break $E_6$ into $SO(10)\tm U(1)$. Such VEVs can be written as $27\tm 27$ matrix as follows
\beqa
<\Phi>^{\bf 650}=\f{\hat{v}_{\bf 650}}{12\sqrt{5}}{\rm diag}(\underbrace{-{5},\cdots,-{5}}_{16},\underbrace{~4,\cdots,~4}_{10},~40)~.
\eeqa
with normalization factor $c=3$.

There are two possible ways to fill the matter contents into $SO(10)\tm U(1)$.
\begin{itemize}

\item $U(1)$ Extension of Ordinary $SO(10)$:

  In this scenario, the Standard Model matter contents can be filled into ${\bf 16_1}$ representation within ${\bf 27}$.
After ${\bf 78}$ dimensional Higgs acquire VEVs, all the sfermions acquire masses
\beqa
 m_{\tl{f}}^2&=&(m_0^{U})^2+{\f{\hat{v}_{\bf 78}}{2\sqrt{6}M_*}}{\beta'}^{\bf 78}(m_0^{N})^2 .
\eeqa Here and in the following sections, we define the universal part for soft sfermion masses \beqa
(m_0^{U})^2=\f{\beta}{M_*^2}F_S^*F_S~,\eeqa
and mass parameters within non-universal part for soft sfermion masses
\beqa
(m_0^{N})^2=\f{1}{2M_*^2}F_T^*F_T~.
\eeqa
After ${\bf 650}$ dimensional Higgs acquire VEVs, all the sfermions acquire masses
\beqa
 m_{\tl{f}}^2&=&(m_0^{U})^2-\f{5 \hat{v}_{\bf 650}}{12\sqrt{5}M_*}{\beta'}^{\bf 650}(m_0^{N})^2 .
\eeqa
\item Flipped SO(10):

  In this scenario, the matter contents are filled as (in notation of SU(5))
 \beqa
 {\bf 16_1}&=&{\bf (10_{\rm Q,D_L^c,N_L^c},~\bar{5}_V,~1_V)}~,\\
 {\bf 10_{-2}}&=&{\bf ( 5_V ,\bar{5}_{\rm U_L^c,L_L} )}~,\\
 {\bf 1_{-4}}&=&{\bf 1_{\rm E_L^c}}~.
 \eeqa

After ${\bf 78}$ dimensional Higgs acquire VEVs, the sfermions acquire masses
\beqa
m_{\tl{Q_L}}^2&=&(m_0^{U})^2+\f{\hat{v}_{\bf 78}}{2\sqrt{6}M_*}{\beta'}^{\bf 78}(m_0^{N})^2 ~,\\
m_{\tl{U_L^C}}^2&=&(m_0^{U})^2-\f{\hat{v}_{\bf 78}}{\sqrt{6}M_*}{\beta'}^{\bf 78}(m_0^{N})^2 ~,\\
m_{\tl{D_L^C}}^2&=&(m_0^{U})^2+\f{\hat{v}_{\bf 78}}{2\sqrt{6}M_*}{\beta'}^{\bf 78}(m_0^{N})^2 ~,\\
m_{\tl{L_L}}^2&=&(m_0^{U})^2-\f{\hat{v}_{\bf 78}}{\sqrt{6}M_*}{\beta'}^{\bf 78}(m_0^{N})^2 ~,\\
m_{\tl{E_L^C}}^2&=&(m_0^{U})^2+2\f{\hat{v}_{\bf 78}}{\sqrt{6}M_*}{\beta'}^{\bf 78}(m_0^{N})^2 .
\eeqa
After ${\bf 650}$ dimensional Higgs acquire VEVs, the sfermions acquire masses
\beqa
m_{\tl{Q_L}}^2&=&(m_0^{U})^2-5\f{\hat{v}_{\bf 650}}{12\sqrt{5}M_*}{\beta'}^{\bf 650}(m_0^{N})^2 ~,\\
m_{\tl{U_L^C}}^2&=&(m_0^{U})^2+\f{\hat{v}_{\bf 650}}{3\sqrt{5}M_*}{\beta'}^{\bf 650}(m_0^{N})^2 ~,\\
m_{\tl{D_L^C}}^2&=&(m_0^{U})^2-5\f{\hat{v}_{\bf 650}}{12\sqrt{5}M_*}{\beta'}^{\bf 650}(m_0^{N})^2 ~,\\
m_{\tl{L_L}}^2&=&(m_0^{U})^2+\f{\hat{v}_{\bf 650}}{3\sqrt{5}M_*}{\beta'}^{\bf 650}(m_0^{N})^2 ~,\\
m_{\tl{E_L^C}}^2&=&(m_0^{U})^2+10\f{\hat{v}_{\bf 650}}{3\sqrt{5}M_*}{\beta'}^{\bf 650}(m_0^{N})^2 .
\eeqa

\end{itemize}

\subsection{$E_6$ To Flipped SU(5) Model}
 There are various symmetry breaking chains in the subsequent $SO(10)\tm U(1)$ breaking.
 There are two possible symmetry breaking chains for $E_6$ to break into flipped SU(5):
\beqa
&& E_6 \ra SO(10)\tm U(1)_1 \ra {\rm flipped} ~SU(5) \tm U(1)_1~,\nn \\
&& E_6 \ra {\rm flipped}~ SO(10) \ra {\rm flipped}~ SU(5)~. \nn
\eeqa
The ${\bf (45,0)}$ components in ${\bf 78}$ and ${\bf 650}$ dimensional representation Higgs of $E_6$ can acquire a VEV which break $E_6$ into $SU(5)\tm U(1)$.
We will not discuss the subsequent breaking chains of ordinary SO(10) because they have already been discussed in \cite{fei1}. Here we concentrate on the breaking of flipped SO(10) into flipped SU(5).

 The ${\bf 78}$ dimensional representation Higgs can acquire Vacuum Expectation Values (VEVs)
which break $E_6$ into $SU(5)\tm U(1)_1 \tm U(1)_2$. Such VEVs can be written as $27\tm 27$ matrix as follows
\beqa
\small
<\Phi>^{\bf 78}_{\bf (45,0)}=\f{v_{\bf 78}}{2\sqrt{10}}{\rm diag}(\underbrace{-1,\cdots,-1}_{10},\underbrace{~3,\cdots,~3}_{5}, -5,\underbrace{~2,\cdots,~2}_{5},\underbrace{-2,\cdots,-2}_{5},~0),\nn\\
\eeqa
\normalsize
with normalization factor $c=3$.
The ${\bf 650}$ dimensional Higgs can also acquire Vacuum Expectation Values (VEVs)
which break $E_6$ into $SU(5)\tm U(1)_1 \tm U(1)_2$. Such VEVs can be written as $27\tm 27$ matrix as follows
\beqa
\small
<\Phi>^{\bf 650}_{\bf (45,0)}=\f{v_{\bf 650}}{4\sqrt{5}}{\rm diag}(\underbrace{~1,\cdots,~1}_{10},\underbrace{-3,\cdots,-3}_{5}, ~5,\underbrace{~4,\cdots,~4}_{5},\underbrace{-4,\cdots,-4}_{5},~0),\nn\\
\eeqa
with normalization factor $c=3$.

After ${\bf (45,0)}$ component of ${\bf 78}$ dimensional Higgs acquire VEVs, the sfermions acquire masses
\beqa
m_{\tl{Q_L}}^2&=&(m_0^{U})^2-\f{v_{\bf 78}}{2\sqrt{10}M_*}{\beta'}^{\bf 78}(m_0^{N})^2 ~,\\
m_{\tl{U_L^C}}^2&=&(m_0^{U})^2-\f{v_{\bf 78}}{\sqrt{10}M_*}{\beta'}^{\bf 78}(m_0^{N})^2 ~,\\
m_{\tl{D_L^C}}^2&=&(m_0^{U})^2-\f{v_{\bf 78}}{2\sqrt{10}M_*}{\beta'}^{\bf 78}(m_0^{N})^2 ~,\\
m_{\tl{L_L}}^2&=&(m_0^{U})^2-\f{v_{\bf 78}}{\sqrt{10}M_*}{\beta'}^{\bf 78}(m_0^{N})^2 ~,\\
m_{\tl{E_L^C}}^2&=&(m_0^{U})^2 ~.
\eeqa

After ${\bf (45,0)}$ component of ${\bf 650}$ dimensional Higgs acquire VEVs, the sfermions acquire masses
\beqa
m_{\tl{Q_L}}^2&=&(m_0^{U})^2+\f{v_{\bf 650}}{4\sqrt{5}M_*}{\beta'}^{\bf 650}(m_0^{N})^2 ~,\\
m_{\tl{U_L^C}}^2&=&(m_0^{U})^2-\f{v_{\bf 650}}{\sqrt{5}M_*}{\beta'}^{\bf 650}(m_0^{N})^2 ~,\\
m_{\tl{D_L^C}}^2&=&(m_0^{U})^2+\f{v_{\bf 650}}{4\sqrt{5}M_*}{\beta'}^{\bf 650}(m_0^{N})^2 ~,\\
m_{\tl{L_L}}^2&=&(m_0^{U})^2-\f{v_{\bf 650}}{\sqrt{5}M_*}{\beta'}^{\bf 650}(m_0^{N})^2~,\\
m_{\tl{E_L^C}}^2&=&(m_0^{U})^2 ~.
\eeqa

\subsection{$E_6$ To $SU(3)_C\tm SU(3)_L\tm SU(3)_R$ Model}
 The fundamental representation ${\bf 27}$ and adjoint representation ${\bf 78}$ of $E_6$ can be decomposed in term of $SU(3)_C\tm SU(3)_L\tm SU(3)_R$
\beqa
\small
{\bf 27}&=&{\bf (~3,~{3},~1)}\oplus {\bf (~\bar{3},~1,~\bar{3})}\oplus {\bf (~1,~\bar{3},~3)}~,\\
{\bf 78}&=&{\bf (~8,~1,~1)}\oplus {\bf (~1, ~8,~1)}\oplus {\bf (~1,~1,~8)}\oplus {\bf (~3,~\bar{3},~\bar{3})}\oplus {\bf (\bar{3},~3,~{3})}.\nn\\
\eeqa
\normalsize

 There are no $SU(3)_C\tm SU(3)_L\tm SU(3)_R$ singlet in decomposition of adjoint Higgs $\Phi({\bf 78})$. So we consider the Vacuum Expectation Values (VEVs) of ${\bf 650}$ dimensional representations which can break $E_6$ into the gauge group $SU(3)_C\tm SU(3)_L\tm SU(3)_R$. There are two singlets in the decomposition of ${\bf 650}$ dimensional representations which we can parameter as $27\tm 27$ matrices.

 The two singlets can be recombined to give one left-right symmetric VEVs which preserve the left-right parity and the other left-right non-symmetric VEVs which breaks the left-right parity.
 The left-right symmetric VEVs can be chosen as
\beqa
\small
<{\bf 650}>_1=\f{v_{\bf 650}}{3\sqrt{2}}{\rm diag}(\underbrace{-2,\cdots,-2}_{9},\underbrace{~1,\cdots,~1}_{9}, \underbrace{~1,\cdots,~1}_{9})~,
\eeqa
\normalsize
with normalization factor $c=3$.
So after ${\bf 650}$ dimensional Higgs acquires such VEVs, we can get the soft supersymmetry breaking mass terms for sfermions
\beqa
\label{33sm-6501}
\small
m_{\tl{Q_L}}^2&=&(m_0^{U})^2+\f{v_{\bf 650}}{3\sqrt{2}M_*}{\beta'}_s^{\bf 650}(m_0^{N})^2 ~,\\
m_{\tl{U_L^C}}^2&=&(m_0^{U})^2+\f{v_{\bf 650}}{3\sqrt{2}M_*}{\beta'}_s^{\bf 650}(m_0^{N})^2 ~,\\
m_{\tl{D_L^C}}^2&=&(m_0^{U})^2+\f{v_{\bf 650}}{3\sqrt{2}M_*}{\beta'}_s^{\bf 650}(m_0^{N})^2 ~,\\
m_{\tl{L_L}}^2&=&(m_0^{U})^2-2\f{v_{\bf 650}}{3\sqrt{2}M_*}{\beta'}_s^{\bf 650}(m_0^{N})^2 ~,\\
m_{\tl{E_L^C}}^2&=&(m_0^{U})^2-2\f{v_{\bf 650}}{3\sqrt{2}M_*}{\beta'}_s^{\bf 650}(m_0^{N})^2 ~.
\eeqa

The other left-right non-symmetric VEVs can be chosen to be
\beqa
<{\bf 650}>_2=\f{\tl{v}_{\bf 650}}{\sqrt{6}}{\rm diag}(\underbrace{~0,\cdots,~0}_{9},\underbrace{~1,\cdots,~1}_{9}, \underbrace{-1,\cdots,-1}_{9})~,
\eeqa
with normalization factor $c=3$.
So after ${\bf 650}$ dimensional Higgs acquires such VEVs, we can get the soft supersymmetry breaking mass terms for sfermions
\beqa
\label{33sm-6502}
m_{\tl{Q_L}}^2&=&(m_0^{U})^2+\f{\tl{v}_{\bf 650}}{\sqrt{6}M_*}{\beta'}_n^{\bf 650}(m_0^{N})^2 ~,\\
m_{\tl{U_L^C}}^2&=&(m_0^{U})^2-\f{\tl{v}_{\bf 650}}{\sqrt{6}M_*}{\beta'}_n^{\bf 650}(m_0^{N})^2 ~,\\
m_{\tl{D_L^C}}^2&=&(m_0^{U})^2-\f{\tl{v}_{\bf 650}}{\sqrt{6}M_*}{\beta'}_n^{\bf 650}(m_0^{N})^2 ~,\\
m_{\tl{L_L}}^2&=&(m_0^{U})^2 ~,\\
m_{\tl{E_L^C}}^2&=&(m_0^{U})^2 ~.
\eeqa
\subsection{$E_6$ To $SU(3)_C\tm SU(2)_L\tm SU(2)_R\tm U(1)_1\tm U(1)_2$ Model}
 This symmetry broken chain can be realized via the VEVs of ${\bf (~1, ~8,~8)}$ components in ${\bf 650}$ dimensional representation
\beqa
\small
<{\bf 650}>=\f{\hat{v}_{\bf 650}}{2\sqrt{3}}{\rm diag}({~1,~1,-2,~1,~1,-2,-2,-2,~4}, \underbrace{~0,\cdots,~0}_{9},\underbrace{~0,\cdots,~0}_{9}),\nn\\
\eeqa\normalsize
with normalization factor $c=3$.
So after ${\bf 650}$ dimensional Higgs acquires such VEVs, we can get the soft supersymmetry breaking mass terms for sfermions
\beqa
m_{\tl{Q_L}}^2&=&(m_0^{U})^2~,\\
m_{\tl{U_L^C}}^2&=&(m_0^{U})^2~,\\
m_{\tl{D_L^C}}^2&=&(m_0^{U})^2~,\\
m_{\tl{L_L}}^2&=&(m_0^{U})^2-2\f{\hat{v}_{\bf 650}}{2\sqrt{3}M_*}{\beta'}^{\bf 650}(m_0^{N})^2 ~,\\
m_{\tl{E_L^C}}^2&=&(m_0^{U})^2-2\f{\hat{v}_{\bf 650}}{2\sqrt{3}M_*}{\beta'}^{\bf 650}(m_0^{N})^2 .
\eeqa
Besides, this symmetry broken chain can also be realized by the VEVs of both $({\bf ~1,~1,~8})$ and $({\bf ~1,~8,~1})$ components of ${\bf 78}$ dimensional representation
\beqa
<{\bf 78}>_1&=&\f{{v}_{\bf 78}}{\sqrt{6}}{\rm diag}(\underbrace{~0,\cdots,~0}_{9},\underbrace{~1,~1,-2}_{3},\underbrace{~0,\cdots,~0}_{9})~,\\
<{\bf 78}>_2&=&\f{\tl{v}_{\bf 78}}{\sqrt{6}}{\rm diag}(\underbrace{~0,\cdots,~0}_{9},\underbrace{~0,\cdots,~0}_{9},\underbrace{~1,~1,-2}_{3})~,
\eeqa
with normalization factor $c=3$.
This symmetry broken chain can also be realized by the VEVs of both $({\bf ~1,~1,~8})$ and $({\bf ~1,~8,~1})$ components of ${\bf 650}$ dimensional representation
\beqa
<{\bf 650}>_1&=&\f{\hat{v}_{\bf 650}^\pr}{\sqrt{6}}{\rm diag}(\underbrace{~1,~1,-2}_{3},\underbrace{~0,\cdots,~0}_{9},\underbrace{~0,\cdots,~0}_{9})~,\\
<{\bf 650}>_2&=&\f{\hat{v}^{\pr\pr}_{\bf 650}}{\sqrt{6}}{\rm diag}(\underbrace{~1,\cdots,~1}_{6},\underbrace{-2,\cdots,-2}_{3},\underbrace{~0,\cdots,~0}_{9},\underbrace{~0,\cdots,~0}_{9}),
\eeqa
with normalization factor $c=3$.
The most general possibilities for $E_6$ breaking into $SU(3)_C\tm SU(2)_L\tm SU(2)_R\tm U(1)_1\tm U(1)_2$ are realized by both the $({\bf ~1,~1,~8})$ VEVs (from ${\bf 78}$ or ${\bf 650}$ dimensional representations) and the $({\bf ~1,~8,~1})$ VEVs (from ${\bf 78}$ or ${\bf 650}$ dimensional representations). Thus the supersymmetry breaking soft mass terms for sfermions
\beqa
m_{\tl{Q_L}}^2&=&(m_0^{U})^2+\f{{v}_{\bf 78}}{\sqrt{6}M_*}{\beta'}^{\bf 78_1}(m_0^{N})^2~,\\
m_{\tl{U_L^C}}^2&=&(m_0^{U})^2+\f{\tl{v}_{\bf 78}}{\sqrt{6}M_*}{\beta'}^{\bf 78_2}(m_0^{N})^2~,\\
m_{\tl{D_L^C}}^2&=&(m_0^{U})^2+\f{\tl{v}_{\bf 78}}{\sqrt{6}M_*}{\beta'}^{\bf 78_2}(m_0^{N})^2~,\\
m_{\tl{L_L}}^2&=&(m_0^{U})^2+\f{\hat{v}_{\bf 650}^\pr}{\sqrt{6}M_*}{\beta'}^{\bf 650_1}(m_0^{N})^2-2\f{\hat{v}_{\bf 650}^{\pr\pr}}{\sqrt{6}M_*}{\beta'}^{\bf 650_2}(m_0^{N})^2 ~,\\
m_{\tl{E_L^C}}^2&=&(m_0^{U})^2-2\f{\hat{v}_{\bf 650}^\pr}{\sqrt{6}M_*}{\beta'}^{\bf 650_1}(m_0^{N})^2+\f{\hat{v}_{\bf 650}^{\pr\pr}}{\sqrt{6}M_*}{\beta'}^{\bf 650_2}(m_0^{N})^2 .
\eeqa
\subsection{$E_6$ To $SU(6)\tm SU(2)$ Model}
$E_6$ GUT can break into $SU(6)\tm SU(2)$ by ${\bf 650}$ dimensional VEVs. According to three different embedding of the standard model matter contents into $SU(6)\times SU(2)$, we investigate three different cases according to the three different choices of SU(2)(namely $SU(2)_X$, $SU(2)_L$ and $SU(2)_R$, respectively).
 The fundamental representation ${\bf 27}$ and adjoint representation ${\bf 78}$ of $E_6$ can be decomposed in term of $SU(6) \tm SU(2)$
\beqa
{\bf 27}&=&{\bf (~\bar{6},~{2})}\oplus {\bf (~15,~1)}~,\\
{\bf 78}&=&{\bf (~35,~1)}\oplus {\bf (~1, ~3)}\oplus {\bf (~20,~2)}~.
\eeqa
The VEVs that break $E_6$ into $SU(6)\tm SU(2)$ can be chosen as
\beqa
<{\bf 650}>=\f{v_{\bf 650}}{6\sqrt{5}}{\rm diag}(\underbrace{-4,\cdots,-4}_{15},\underbrace{~5,\cdots,~5}_{12})~,
\eeqa
with normalization factor $c=3$.

Then we have three possibilities relating to different filling of the standard model matter contents
\begin{itemize}
\item $E_6\ra SU(6)\tm SU(2)_X\ra SU(5)\tm U(1)\tm SU(2)_X$:

 After ${\bf 650}$ dimensional Higgs acquires VEVs, the supersymmetry breaking soft mass terms for sfermions
\beqa
m_{\tl{Q_L}}^2&=&(m_0^{U})^2-\f{4{v}_{\bf 650}}{6\sqrt{5}M_*}{\beta'}^{\bf 650}(m_0^{N})^2~,\\
m_{\tl{U_L^C}}^2&=&(m_0^{U})^2-\f{4{v}_{\bf 650}}{6\sqrt{5}M_*}{\beta'}^{\bf 650}(m_0^{N})^2~,\\
m_{\tl{D_L^C}}^2&=&(m_0^{U})^2+\f{5{v}_{\bf 650}}{6\sqrt{5}M_*}{\beta'}^{\bf 650}(m_0^{N})^2~~,\\
m_{\tl{L_L}}^2&=&(m_0^{U})^2+\f{5{v}_{\bf 650}}{6\sqrt{5}M_*}{\beta'}^{\bf 650}(m_0^{N})^2~ ~,\\
m_{\tl{E_L^C}}^2&=&(m_0^{U})^2-\f{4{v}_{\bf 650}}{6\sqrt{5}M_*}{\beta'}^{\bf 650}(m_0^{N})^2~.
\eeqa

\item $E_6\ra SU(6)\tm SU(2)_L\ra SU(4)_c\tm SU(2)_L\tm SU(2)_R \tm U(1)_1$:

After ${\bf 650}$ dimensional Higgs acquires VEVs, the supersymmetry broken soft mass terms for sfermions
\beqa
m_{\tl{Q_L}}^2&=&(m_0^{U})^2+\f{5{v}_{\bf 650}}{6\sqrt{5}M_*}{\beta'}^{\bf 650}(m_0^{N})^2~,\\
m_{\tl{U_L^C}}^2&=&(m_0^{U})^2-\f{4{v}_{\bf 650}}{6\sqrt{5}M_*}{\beta'}^{\bf 650}(m_0^{N})^2~,\\
m_{\tl{D_L^C}}^2&=&(m_0^{U})^2-\f{4{v}_{\bf 650}}{6\sqrt{5}M_*}{\beta'}^{\bf 650}(m_0^{N})^2~~,\\
m_{\tl{L_L}}^2&=&(m_0^{U})^2+\f{5{v}_{\bf 650}}{6\sqrt{5}M_*}{\beta'}^{\bf 650}(m_0^{N})^2~ ~,\\
m_{\tl{E_L^C}}^2&=&(m_0^{U})^2-\f{4{v}_{\bf 650}}{6\sqrt{5}M_*}{\beta'}^{\bf 650}(m_0^{N})^2~.
\eeqa

\item $E_6\ra SU(6)\tm SU(2)_R \ra SU(4)_c\tm SU(2)_L\tm SU(2)_R \tm U(1)_2$:

After ${\bf 650}$ dimensional Higgs acquires VEVs, the supersymmetry broken soft mass terms for sfermions
\beqa
m_{\tl{Q_L}}^2&=&(m_0^{U})^2-\f{4{v}_{\bf 650}}{6\sqrt{5}M_*}{\beta'}^{\bf 650}(m_0^{N})^2~,\\
m_{\tl{U_L^C}}^2&=&(m_0^{U})^2+\f{5{v}_{\bf 650}}{6\sqrt{5}M_*}{\beta'}^{\bf 650}(m_0^{N})^2~,\\
m_{\tl{D_L^C}}^2&=&(m_0^{U})^2+\f{5{v}_{\bf 650}}{6\sqrt{5}M_*}{\beta'}^{\bf 650}(m_0^{N})^2~~,\\
m_{\tl{L_L}}^2&=&(m_0^{U})^2-\f{4{v}_{\bf 650}}{6\sqrt{5}M_*}{\beta'}^{\bf 650}(m_0^{N})^2~ ~,\\
m_{\tl{E_L^C}}^2&=&(m_0^{U})^2+\f{5{v}_{\bf 650}}{6\sqrt{5}M_*}{\beta'}^{\bf 650}(m_0^{N})^2~.
\eeqa
\end{itemize}

\subsection{$E_6$ To $SU(4)_c\tm SU(2)_L \tm SU(2)_R \tm U(1)$ Model}

There are two possible symmetry broken chains for $E_6$ breaking into Pati-Salam model.
One symmetry breaking chain is
\beqa
E_6\ra SO(10)\tm U(1)\ra SU(4)_{c}\tm SU(2)_L\tm SU(2)_R\tm U(1)~,
\eeqa
the other symmetry breaking chain is
\beqa
E_6\ra SU(6)\tm SU(2)_{L,R}\ra SU(4)_{c}\tm SU(2)\tm SU(2)_R\tm U(1)~.
\eeqa

In this subsection, we concentrate on the second one.
Such breaking can be realized via the VEVs of ${\bf 78}$ and ${\bf 650}$ dimensional representations.

The ${\bf (~35,~1)}$ component VEVs of the ${\bf 78}$ dimensional representation that break gauge group $SU(6)\tm SU(2)_1$  into $SU(4)\tm SU(2)_1\tm SU(2)_2 \tm U(1)$ reads
\beqa\small
\langle{\bf 78}\rangle_{\bf (35,1)}=\f{v_{\bf 78}}{2\sqrt{6}}{\rm diag}(\underbrace{~1,~1,~1,~1,-2,-2}_{2},\underbrace{~2,~\cdots,~2}_{6},\underbrace{-1,~\cdots,-1}_{8},-4),\nn\\
\eeqa\normalsize
with normalization factor $c=3$.
The breaking of gauge group $SU(6)\tm SU(2)_1$  into $SU(4)\tm SU(2)_1\tm SU(2)_2 \tm U(1)$ can also be realized by both the ${\bf (35,1)}$ and the ${\bf (189,1)}$ component VEVs of ${\bf 650}$ dimensional representation
\beqa\small
\langle{\bf 650}\rangle_{\bf (35,1)}&=&\f{{v'}_{\bf 650}}{2\sqrt{3}}{\rm diag}(\underbrace{~1,~1,~1,~1,-2,-2}_{2},\underbrace{-1,~\cdots,-1}_{6},\underbrace{~\f{1}{2},~\cdots,~\f{1}{2}}_{8},~2),\nn\\
\langle{\bf 650}\rangle_{\bf (189,1)}&=&\f{\tl{v'}_{\bf 650}}{4\sqrt{5}}{\rm diag}(\underbrace{~0,\cdots,~0}_{12},\underbrace{-2,~\cdots,-2}_{6},\underbrace{~3,~\cdots,~3}_{8},-12),\nn\\
\eeqa\normalsize
with normalization factor $c=3$.

\begin{itemize}

\item $E_6\ra SU(6)\tm SU(2)_L\ra SU(4)_c\tm SU(2)_L\tm SU(2)_R \tm U(1)_1$:

After the ${\bf (~35,~1)}$ component of ${\bf 78}$ dimensional Higgs acquires VEVs, we can get the soft supersymmetry breaking mass terms for sfermions
\beqa
m_{\tl{Q_L}}^2&=&(m_0^{U})^2+\f{{v}_{\bf 78}}{2\sqrt{6}M_*}{\beta'}^{\bf 78}(m_0^{N})^2~,\\
m_{\tl{U_L^C}}^2&=&(m_0^{U})^2-\f{{v}_{\bf 78}}{2\sqrt{6}M_*}{\beta'}^{\bf 78}(m_0^{N})^2~,\\
m_{\tl{D_L^C}}^2&=&(m_0^{U})^2-\f{{v}_{\bf 78}}{2\sqrt{6}M_*}{\beta'}^{\bf 78}(m_0^{N})^2~~,\\
m_{\tl{L_L}}^2&=&(m_0^{U})^2+\f{{v}_{\bf 78}}{2\sqrt{6}M_*}{\beta'}^{\bf 78}(m_0^{N})^2~ ~,\\
m_{\tl{E_L^C}}^2&=&(m_0^{U})^2-\f{{v}_{\bf 78}}{2\sqrt{6}M_*}{\beta'}^{\bf 78}(m_0^{N})^2~.
\eeqa
After the ${\bf (~35,~1)}$ component of ${\bf 650}$ dimensional Higgs acquires VEVs, we can get the soft supersymmetry breaking mass terms for sfermions
\beqa
m_{\tl{Q_L}}^2&=&(m_0^{U})^2+\f{{v'}_{\bf 650}}{2\sqrt{3}M_*}{\beta'}^{\bf 650}(m_0^{N})^2~,\\
m_{\tl{U_L^C}}^2&=&(m_0^{U})^2+\f{{v'}_{\bf 650}}{4\sqrt{3}M_*}{\beta'}^{\bf 650}(m_0^{N})^2~,\\
m_{\tl{D_L^C}}^2&=&(m_0^{U})^2+\f{{v'}_{\bf 650}}{4\sqrt{3}M_*}{\beta'}^{\bf 650}(m_0^{N})^2~~,\\
m_{\tl{L_L}}^2&=&(m_0^{U})^2+\f{{v'}_{\bf 650}}{2\sqrt{3}M_*}{\beta'}^{\bf 650}(m_0^{N})^2~ ~,\\
m_{\tl{E_L^C}}^2&=&(m_0^{U})^2+\f{{v'}_{\bf 650}}{4\sqrt{3}M_*}{\beta'}^{\bf 650}(m_0^{N})^2~ .
\eeqa
After the ${\bf (~189,~1)}$ component of ${\bf 650}$ dimensional Higgs acquires VEVs, we can get the soft supersymmetry breaking mass terms for sfermions
\beqa
m_{\tl{Q_L}}^2&=&(m_0^{U})^2~,\\
m_{\tl{U_L^C}}^2&=&(m_0^{U})^2+\f{3\tl{v'}_{\bf 650}}{4\sqrt{5}M_*}{\beta'}^{\bf 650}(m_0^{N})^2~,\\
m_{\tl{D_L^C}}^2&=&(m_0^{U})^2+\f{3\tl{v'}_{\bf 650}}{4\sqrt{5}M_*}{\beta'}^{\bf 650}(m_0^{N})^2~~,\\
m_{\tl{L_L}}^2&=&(m_0^{U})^2~,\\
m_{\tl{E_L^C}}^2&=&(m_0^{U})^2+\f{3\tl{v'}_{\bf 650}}{4\sqrt{5}M_*}{\beta'}^{\bf 650}(m_0^{N})^2~.
\eeqa

\item $E_6\ra SU(6)\tm SU(2)_R \ra SU(4)_c\tm SU(2)_L\tm SU(2)_R \tm U(1)_2$:

After the ${\bf (~35,~1)}$ component of ${\bf 78}$ dimensional Higgs acquires VEVs, we can get the soft supersymmetry breaking mass terms for sfermions
\beqa
m_{\tl{Q_L}}^2&=&(m_0^{U})^2-\f{{v}_{\bf 78}}{2\sqrt{6}M_*}{\beta'}^{\bf 78}(m_0^{N})^2~,\\
m_{\tl{U_L^C}}^2&=&(m_0^{U})^2+\f{{v}_{\bf 78}}{2\sqrt{6}M_*}{\beta'}^{\bf 78}(m_0^{N})^2~,\\
m_{\tl{D_L^C}}^2&=&(m_0^{U})^2+\f{{v}_{\bf 78}}{2\sqrt{6}M_*}{\beta'}^{\bf 78}(m_0^{N})^2~~,\\
m_{\tl{L_L}}^2&=&(m_0^{U})^2-\f{{v}_{\bf 78}}{2\sqrt{6}M_*}{\beta'}^{\bf 78}(m_0^{N})^2~ ~,\\
m_{\tl{E_L^C}}^2&=&(m_0^{U})^2+\f{{v}_{\bf 78}}{2\sqrt{6}M_*}{\beta'}^{\bf 78}(m_0^{N})^2.
\eeqa

After the ${\bf (~35,~1)}$ component of ${\bf 650}$ dimensional Higgs acquires VEVs, we can get the soft supersymmetry breaking mass terms for sfermions
\beqa
m_{\tl{Q_L}}^2&=&(m_0^{U})^2+\f{{v'}_{\bf 650}}{4\sqrt{3}M_*}{\beta'}^{\bf 650}(m_0^{N})^2~,\\
m_{\tl{U_L^C}}^2&=&(m_0^{U})^2+\f{{v'}_{\bf 650}}{2\sqrt{3}M_*}{\beta'}^{\bf 650}(m_0^{N})^2~,\\
m_{\tl{D_L^C}}^2&=&(m_0^{U})^2+\f{{v'}_{\bf 650}}{2\sqrt{3}M_*}{\beta'}^{\bf 650}(m_0^{N})^2~~,\\
m_{\tl{L_L}}^2&=&(m_0^{U})^2+\f{{v'}_{\bf 650}}{4\sqrt{3}M_*}{\beta'}^{\bf 650}(m_0^{N})^2~ ~,\\
m_{\tl{E_L^C}}^2&=&(m_0^{U})^2+\f{{v'}_{\bf 650}}{2\sqrt{3}M_*}{\beta'}^{\bf 650}(m_0^{N})^2.
\eeqa

After the ${\bf (~189,~1)}$ component of ${\bf 650}$ dimensional Higgs acquires VEVs, we can get the soft supersymmetry breaking mass terms for sfermions
\beqa
m_{\tl{Q_L}}^2&=&(m_0^{U})^2+\f{3\tl{v'}_{\bf 650}}{4\sqrt{5}M_*}{\beta'}^{\bf 650}(m_0^{N})^2~,\\
m_{\tl{U_L^C}}^2&=&(m_0^{U})^2~,\\
m_{\tl{D_L^C}}^2&=&(m_0^{U})^2~,\\
m_{\tl{L_L}}^2&=&(m_0^{U})^2+\f{3\tl{v'}_{\bf 650}}{4\sqrt{5}M_*}{\beta'}^{\bf 650}(m_0^{N})^2~,\\
m_{\tl{E_L^C}}^2&=&(m_0^{U})^2 ~,
\eeqa

\end{itemize}

\subsection{$E_6$ To $SU(5)\tm U(1) \tm SU(2)_X$ Model}
The breaking of $E_6$ into $SU(5)\tm U(1) \tm SU(2)_X$ can be realized via the VEVs of ${\bf 78}$ and ${\bf 650}$ dimensional representations.

The ${\bf (~35,~1)}$ component VEVs of the ${\bf 78}$ dimensional representation that break $SU(6)\tm SU(2)_X$ to $SU(5)\tm U(1) \tm SU(2)_X$ reads
\beqa\small
<{\bf 78}>_{\bf (35,1)}=\f{\hat{v}_{\bf 78}}{2\sqrt{15}}{\rm diag}(\underbrace{~1,~1,~1,~1,~1,-5}_{2},\underbrace{~2,~\cdots,~2}_{10},\underbrace{-4,~\cdots,-4}_{5}),
\eeqa\normalsize
with normalization factor $c=3$.
The ${\bf (~35,~1)}$ component VEVs of the ${\bf 650}$ dimensional representation that break gauge group $SU(6)\tm SU(2)_X$ into group $SU(5)\tm U(1) \tm SU(2)_X$ reads
\beqa\small
<{\bf 650}>_{\bf (35,1)}=\f{\hat{v}_{\bf 650}}{\sqrt{30}}{\rm diag}(\underbrace{~1,~1,~1,~1,~1,-5}_{2},\underbrace{-1,~\cdots,-1}_{10},\underbrace{~2,~\cdots,~2}_{5})~,
\eeqa\normalsize
with normalization factor $c=3$.
After the ${\bf (~35,~1)}$ component of ${\bf 78}$ dimensional Higgs acquires VEVs, we can get the soft supersymmetry breaking mass terms for sfermions
\beqa
m_{\tl{Q_L}}^2&=&(m_0^{U})^2+\f{\hat{v}_{\bf 78}}{\sqrt{15}M_*}{\beta'}^{\bf 78}(m_0^{N})^2~,\\
m_{\tl{U_L^C}}^2&=&(m_0^{U})^2+\f{\hat{v}_{\bf 78}}{\sqrt{15}M_*}{\beta'}^{\bf 78}(m_0^{N})^2~,\\
m_{\tl{D_L^C}}^2&=&(m_0^{U})^2+\f{\hat{v}_{\bf 78}}{2\sqrt{15}M_*}{\beta'}^{\bf 78}(m_0^{N})^2~~,\\
m_{\tl{L_L}}^2&=&(m_0^{U})^2+\f{\hat{v}_{\bf 78}}{2\sqrt{15}M_*}{\beta'}^{\bf 78}(m_0^{N})^2~ ~,\\
m_{\tl{E_L^C}}^2&=&(m_0^{U})^2+\f{\hat{v}_{\bf 78}}{\sqrt{15}M_*}{\beta'}^{\bf 78}(m_0^{N})^2~.
\eeqa
After the ${\bf (~35,~1)}$ component of ${\bf 650}$ dimensional Higgs acquires VEVs, we can get the soft supersymmetry breaking mass terms for sfermions
\beqa
m_{\tl{Q_L}}^2&=&(m_0^{U})^2-\f{\hat{v}_{\bf 650}}{\sqrt{30}M_*}{\beta'}^{\bf 650}(m_0^{N})^2~,\\
m_{\tl{U_L^C}}^2&=&(m_0^{U})^2-\f{\hat{v}_{\bf 650}}{\sqrt{30}M_*}{\beta'}^{\bf 650}(m_0^{N})^2~,\\
m_{\tl{D_L^C}}^2&=&(m_0^{U})^2+\f{\hat{v}_{\bf 650}}{\sqrt{30}M_*}{\beta'}^{\bf 650}(m_0^{N})^2~,\\
m_{\tl{L_L}}^2&=&(m_0^{U})^2+\f{\hat{v}_{\bf 650}}{\sqrt{30}M_*}{\beta'}^{\bf 650}(m_0^{N})^2 ~,\\
m_{\tl{E_L^C}}^2&=&(m_0^{U})^2-\f{\hat{v}_{\bf 650}}{\sqrt{30}M_*}{\beta'}^{\bf 650}(m_0^{N})^2~.
\eeqa

\section{MSSM Superpotential and Soft Trilinear Terms in $E_6$ SUSY GUT}
\label{sec-3}
To get new contributions to MSSM superpotential and soft trilinear terms from higher-dimensional operators, we need to consider the group tensor production decomposition for the
Yukawa coupling \cite{Slansky:1981yr}\beqa\small {\bf 27}_m\otimes{\bf 27}_m\otimes{\bf
27}_H&=&({\bf \overline{27}_s}\oplus{\bf \overline{351}_a}\oplus{\bf \overline{351}_s^\pr})\otimes {\bf
27}\nn\\&=&({\bf 1}\oplus{\bf 78}\oplus{\bf 650})\oplus({\bf 78}\oplus{\bf
650}\oplus{\bf 2925}\oplus{\bf
\overline{5824}})\nn\\&\oplus& ({\bf 650}\oplus{\bf
3003}\oplus{\bf \overline{5824}})~. \eeqa\normalsize

We consider in this paper the effect of ${\bf 78,650}$ dimensional representation Higgs to superpotential and trilinear terms.
  For ${\bf 78}$ dimensional representation Higgs fields, we consider the following non-renormalizable superpotential
\beqa\small
W&\supset& \f{1}{M_*}\left[\f{}{}\.h_{ij}^a[({\bf 27}_m^i\otimes{\bf 27}_m^j)^a_{\bf \overline{351}}\otimes{\bf 78}]\otimes{\bf 27}_H\nn\\
&&+(h_{ij}^\pr)^s({\bf 27}^i_m\otimes{\bf 27}_m^j)^s_{\bf \overline{27}}\otimes({\bf 27}_H\otimes{\bf 78})\left.\f{}{}\] \nn\\
&+&\al\f{T}{M_*^2}\left[\f{}{}\.y_{ij}^a[({\bf 27}_m^i\otimes{\bf 27}_m^j)^a_{\bf \overline{351}}\otimes{\bf 78}]\otimes{\bf 27}_H\nn\\
&&+ (y_{ij}^\pr)^s({\bf 27}^i_m\otimes{\bf 27}_m^j)^s_{\bf \overline{27}}\otimes({\bf 27}_H\otimes{\bf 78})\left.\f{}{}\right]~.
\eeqa\normalsize
 The two terms corresponds to two linearly independent contraction methods in the group production. The superscripts $'s'$ ( or $'a'$) indicates that the coefficients are symmetric (or antisymmetric) with respect to the family $'ij'$ index.

For ${\bf 650}$ dimensional representation Higgs fields, we consider the following non-renormalizable superpotential
\beqa\scriptsize
&&W\supseteq\f{1}{M_*}\left[\f{}{}h_{ij}^a[({\bf 27}_m^i\otimes{\bf 27}_m^j)^a_{\bf \overline{351}}\otimes{\bf 650}]\otimes{\bf 27}_H\.\nn\\&&+h_{ij}^{\pr s}[({\bf 27}^i_m\otimes{\bf 27}_m^j)^s_{\bf \overline{351}^\pr}\otimes{\bf 650}]\otimes{\bf 27}_H+\left.(h_{ij}^{\pr\pr})^s({\bf 27}_m^i\otimes{\bf 27}_m^j)^s_{\bf \overline{27}}\otimes({\bf 27}_H\otimes{\bf 650}) \f{}{}\right]\nn\\
&&~~~+\al\f{T}{M_*^2}\left[\f{}{}y_{ij}^a[({\bf 27}_m^i\otimes{\bf 27}_m^j)^a_{\bf \overline{351}}\otimes{\bf 650}]\otimes{\bf 27}_H\.\nn\\&&+ y_{ij}^{\pr s}[({\bf 27}^i_m\otimes{\bf 27}_m^j)^s_{\bf \overline{351}^\pr}\otimes{\bf 650}]\otimes{\bf 27}_H+(y_{ij}^{\pr\pr})^s({\bf 27}_m^i\otimes{\bf 27}_m^j)^s_{\bf \overline{27}}\otimes({\bf 27}_H\otimes{\bf 650})\left. \right].\nn\\
\eeqa\normalsize
The three terms corresponds to three linearly independent contraction methods in the group production. The superscripts $'s'$ ( or $'a'$) indicates that the coefficients are symmetric (or antisymmetric) with respect to the family $'ij'$ index.

\subsection{$E_6$ To $SO(10)\tm U(1)$ Model}
 The ${\bf 78}$ dimensional representation Higgs can acquire Vacuum Expectation Values (VEVs)
which break $E_6$ into $SO(10)\tm U(1)$. Such VEVs can be written as $27\tm 27$ matrix as follows
\beqa\small
<\Phi>^{\bf 78}=\f{\hat{v}_{\bf 78}}{2\sqrt{6}}{\rm diag}(\underbrace{~1,\cdots,~1}_{16},\underbrace{-2,\cdots,-2}_{10},~4)~,
\eeqa\normalsize
with normalization factor $c=3$.
The ${\bf 650}$ dimensional Higgs can also acquire Vacuum Expectation Values
which break $E_6$ into $SO(10)\tm U(1)$. Such VEVs can be written as $27\tm 27$ matrix as follows
\beqa\small
<\Phi>^{\bf 650}=\f{\hat{v}_{\bf 650}}{12\sqrt{5}}{\rm diag}(\underbrace{-{5},\cdots,-{5}}_{16},\underbrace{~4,\cdots,~4}_{10},~40)~,
\eeqa\normalsize
with normalization factor $c=3$.
\begin{itemize}
\item U(1) Extension of Ordinary SO(10):

  The gauge invariant Yukawa coupling in $E_6$ GUT have the form
  \beqa
  W&\supset& \sum\limits_{i,j=1}^3y_{ij} {\bf 27}^i{\bf 27}^j{\bf 27}_h\supset\sum\limits_{i,j=1}^3 y_{ij} {\bf 16}^i {\bf 16}^j{\bf 10}_H \nn~\\
   &\supset& \sum\limits_{i,j=1}^32y_{ij}^s \[Q_L^i(U_L^c)^j h_u+ Q_L^i(D_L^c)^j h_d+ L_L^i(E_L^c)^j h_d+ L_L^i(N_L^c)^j h_u\].\nn\\
  \eeqa
 After the ${\bf(45,~1)}$ component of ${\bf 78}$ dimensional Higgs acquire VEVs which is denoted by $\langle{\bf 78}\rangle_{\bf (45,1)}$, the new contributions to superpotential
\beqa W  &\supset &\f{\hat{v}_{\bf 78}}{2\sqrt{6}M_*} \sum\limits_{i,j=1}^3\[ h_{ij}^a{\bf 16}_i {\bf 16}_j{\bf 10}_H
-2 h_{ij}^{\pr s}{\bf 16}_i {\bf 16}_j{\bf 10}_H \f{}{}\right],\nn\\
&\supset &\f{\hat{v}_{\bf 78}}{2\sqrt{6}M_*} \sum\limits_{i,j=1}^3\[-2h_{ij}^{\pr s} \{2Q_L^i(U_L^c)^jH_u+2Q_L^i(D_L^c)^jH_d\.\nn\\&&~~~~~~~~~~~~~+ 2L_L^i(E_L^c)^jH_d+ 2L_L^i(N_L^c)^jH_u\}]\left.\f{}{}\right],
\eeqa
while the new contributions to supersymmetry breaking soft trilinear terms
 \beqa
 -{\cal L} &\supset& \al^\pr\f{\hat{v}_{\bf 78}F_T}{2\sqrt{6}M_*^2}\sum\limits_{i,j=1}^3\[ -2y_{ij}^{\pr s} \{2\tl{Q}_L^i(\tl{U}_L^c)^jH_u+2\tl{Q}_L^i(\tl{D}_L^c)^jH_d \.\nn\\&&~~~~~~~~~~~~~~~+ 2\tl{L}_L^i(\tl{E}_L^c)^jH_d+ 2\tl{L}_L^i(\tl{N}_L^c)^jH_u\}\left.\f{}{}\right]. \eeqa

After ${\bf (~45,~1)}$ component of ${\bf 650}$ dimensional Higgs acquire VEVs which is denoted by $\langle{\bf 650}\rangle_{\bf (45,1)}$, the new contributions to superpotential
\beqa W  &\supset &\f{\hat{v}_{\bf 650}}{12\sqrt{5}M_*} \sum\limits_{i,j=1}^3\[ -5h_{ij}^a{\bf 16}_i {\bf 16}_j{\bf 10}_H
-5h_{ij}^{\pr s}{\bf 16}_i {\bf 16}_j{\bf 10}_H\.\nn\\&&~~~~~~~~~~~~~~~~~~~~~~~+4 h_{ij}^{\pr\pr s}{\bf 16}_i {\bf 16}_j{\bf 10}_H \left.\f{}{}\right],\nn\\
&\supset &\f{\hat{v}_{\bf 650}}{12\sqrt{5}M_*} \sum\limits_{i,j=1}^3\[\f{}{}2(-5h_{ij}^{\pr s}+4 h_{ij}^{\pr\pr s}) \{Q_L^i(U_L^c)^jH_u+Q_L^i(D_L^c)^jH_d \.\nn\\&&~~~~~~~~~~~~~~+ L_L^i(E_L^c)^jH_d+ L_L^i(N_L^c)^jH_u\}\left.\f{}{}\right],
\eeqa
while the new contributions to supersymmetry breaking soft trilinear terms
 \beqa
 -{\cal L} &\supset& \al^\pr\f{\hat{v}_{\bf 650}F_T}{2\sqrt{6}M_*^2}\sum\limits_{i,j=1}^3\[2(-5y_{ij}^{\pr s}+4y_{ij}^{\pr\pr s} ) \{\tl{Q}_L^i(\tl{U}_L^c)^jH_u+\tl{Q}_L^i(\tl{D}_L^c)^jH_d \.\nn\\&&~~~~~~~~~~~~~~~~~~+ \tl{L}_L^i(\tl{E}_L^c)^jH_d+ \tl{L}_L^i(\tl{N}_L^c)^jH_u\}\left.\f{}{}\right].\eeqa

\item Flipped SO(10):

 The gauge invariant Yukawa coupling in $E_6$ GUT have the form
  \beqa
  W&\supset& \sum\limits_{i,j=1}^3y_{ij} {\bf 27}^i{\bf 27}^j{\bf 27}_h\nn\\&\supset&\sum\limits_{i,j=1}^3 y_{ij} {\bf 16}^i_m {\bf 16}^j_m{\bf 10}_H
  +2y_{ij}^s {\bf 16}^i_m {\bf 10}^j_m{\bf 16}_H+2y_{ij}^s{\bf 10}_m{\bf 1}_m{\bf 10}_H\nn~\\
   &\supset& \sum\limits_{i,j=1}^32y_{ij}^s \[Q_L^i(U_L^c)^j h_u+ Q_L^i(D_L^c)^j h_d+ L_L^i(E_L^c)^j h_d+ L_L^i(N_L^c)^j h_u\].\nn\\
  \eeqa
 After the ${\bf (~45,~1)}$ component of ${\bf 78}$ dimensional Higgs acquire VEVs which is denoted by $\langle{\bf 78}\rangle_{\bf (45,1)}$, the new contributions to superpotential
\beqa W  &\supset &\f{\hat{v}_{\bf 78}}{2\sqrt{6}M_*} \sum\limits_{i,j=1}^3\[\f{}{} h_{ij}^a{\bf 16}_i {\bf 16}_j{\bf 10}_H+3h_{ij}^a{\bf 16}^i_m {\bf 10}^j_m{\bf 16}_H\.\nn\\&&~~~-6h_{ij}^a{\bf 10}_m{\bf 1}_m{\bf 10}_H-2 h_{ij}^{\pr s}{\bf 16}_i {\bf 16}_j{\bf 10}_H+2h_{ij}^{\pr s}{\bf 16}^i_m {\bf 10}^j_m{\bf 16}_H\nn\\&&~~~~~~~~~~~~~~-4h_{ij}^{\pr s}{\bf 10}_m{\bf 1}_m{\bf 10}_H\left.\f{}{}\right],\nn\\
&\supset &\f{\hat{v}_{\bf 78}}{2\sqrt{6}M_*} \sum\limits_{i,j=1}^3\[\f{}{}-4h_{ij}^{\pr s}Q_L^i(D_L^c)^jH_d+(3h_{ij}^a+2h_{ij}^{\pr s}) \{Q_L^i(U_L^c)^jH_u\.\nn\\&&~~~~~~~~~~~~~+ L_L^i(E_L^c)^jH_d\}+(-6h_{ij}^a-4h_{ij}^{\pr s})L_L^i(N_L^c)^jH_u\left.\f{}{}\right],\nn\\
\eeqa
while the new contributions to supersymmetry breaking soft trilinear terms
 \beqa
 -{\cal L} &\supset& \al'\f{\hat{v}_{\bf 78}F_T}{2\sqrt{6}M_*^2}\sum\limits_{i,j=1}^3\[ -4y_{ij}^{\pr s}\tl{Q}_L^i(\tl{D}_L^c)^jH_d+ (3y_{ij}^a+2y_{ij}^{\pr s})\{\tl{Q}_L^i(\tl{U}_L^c)^jH_u\.\nn\\&&~~~~~~~+ \tl{L}_L^i(\tl{E}_L^c)^jH_d\}+ (-6y_{ij}^a-4y_{ij}^{\pr s})\tl{L}_L^i(\tl{N}_L^c)^jH_u\}\left.\f{}{}\right]. \eeqa

After ${\bf (~45,~1)}$ component of ${\bf 650}$ dimensional Higgs acquire VEVs which is denoted by $\langle{\bf 650}\rangle_{\bf (45,1)}$, the new contributions to superpotential
\beqa\scriptsize W  &\supset &\f{\hat{v}_{\bf 650}}{12\sqrt{5}M_*} \sum\limits_{i,j=1}^3\[\f{}{}44h_{ij}^{\pr s}{\bf 10}_m{\bf 1}_m{\bf 10}_H -5h_{ij}^a{\bf 16}_i {\bf 16}_j{\bf 10}_H\.\nn\\&&-9h_{ij}^a{\bf 16}^i_m {\bf 10}^j_m{\bf 16}_H-5 h_{ij}^{\pr s}{\bf 16}_i {\bf 16}_j{\bf 10}_H-h_{ij}^{\pr s}{\bf 16}^i_m {\bf 10}^j_m{\bf 16}_H\nn\\
&&-36h_{ij}^a{\bf 10}_m{\bf 1}_m{\bf 10}_H+4 h_{ij}^{\pr\pr s}{\bf 16}_i {\bf 16}_j{\bf 10}_H-10h_{ij}^{\pr\pr s}{\bf 16}^i_m {\bf 10}^j_m{\bf 16}_H\nn\\
&&~~~~~~~~~~~~~~~~~~+8h_{ij}^{\pr\pr s}{\bf 10}_m{\bf 1}_m{\bf 10}_H\left.\f{}{}\right],\nn\\
&\supset &\f{\hat{v}_{\bf 650}}{12\sqrt{5}M_*} \sum\limits_{i,j=1}^3\[(-10h_{ij}^{\pr s}+8h_{ij}^{\pr\pr s})Q_L^i(D_L^c)^jH_d\.\nn\\&&~~~~~~~~~~+(-9h_{ij}^a-h_{ij}^{\pr s}-10h_{ij}^{\pr\pr s})\{Q_L^i(U_L^c)^jH_u+ L_L^i(E_L^c)^jH_d\}\nn\\&&~~~~~~~~~~~+(-36h_{ij}^a+44h_{ij}^{\pr s}+8h_{ij}^{\pr\pr s})L_L^i(N_L^c)^jH_u\left.\f{}{}\right],
\eeqa\normalsize
while the new contributions to supersymmetry breaking soft trilinear terms
 \beqa
 -{\cal L} &\supset& \al'\f{v_{\bf 650}F_T}{12\sqrt{5}M_*^2}\sum\limits_{i,j=1}^3\[ (-10y_{ij}^{\pr s}+8y_{ij}^{\pr\pr s})\tl{Q}_L^i(\tl{D}_L^c)^jH_d\.\nn\\&&~~~~~~+ (-9y_{ij}^a-y_{ij}^{\pr s}-10y_{ij}^{\pr\pr s})\{\tl{Q}_L^i(\tl{U}_L^c)^jH_u+ \tl{L}_L^i(\tl{E}_L^c)^jH_d\}\nn\\&&~~~~~~+ (-36y_{ij}^a+44y_{ij}^{\pr s}+8y_{ij}^{\pr\pr s})\tl{L}_L^i(\tl{N}_L^c)^jH_u\}\left.\f{}{}\right]. \eeqa

\end{itemize}

\subsection{$E_6$ To Flipped SU(5) Model}
As before, we will not discuss new contributions to the trilinear terms from subsequent breaking chains of ordinary SO(10) because they have already
been discussed in our previous works \cite{fei1}. Here we concentrate on the breaking of flipped SO(10) into flipped SU(5).

 The ${\bf 78}$ dimensional representation Higgs can acquire Vacuum Expectation Values
which break $E_6$ into $SU(5)\tm U(1)_1 \tm U(1)_2$. Such VEVs can be written as $27\tm 27$ matrix as follows
\beqa
\small
<\Phi>^{\bf 78}_{\bf (45,0)}=\f{v_{\bf 78}}{2\sqrt{10}}{\rm diag}(\underbrace{-1,\cdots,-1}_{10},\underbrace{~3,\cdots,~3}_{5}, -5,\underbrace{~2,\cdots,~2}_{5},\underbrace{-2,\cdots,-2}_{5},~0),\nn\\
\eeqa
\normalsize
with normalization factor $c=3$.
The ${\bf 650}$ dimensional Higgs can also acquire Vacuum Expectation Values
which break $E_6$ gauge group into its subgroup $SU(5)\tm U(1)_1 \tm U(1)_2$. Such VEVs can be written as $27\tm 27$ matrix as follows
\beqa
\small
<\Phi>^{\bf 650}_{\bf (45,0)}=\f{v_{\bf 650}}{4\sqrt{5}}{\rm diag}(\underbrace{~1,\cdots,~1}_{10},\underbrace{-3,\cdots,-3}_{5}, ~5,\underbrace{~4,\cdots,~4}_{5},\underbrace{-4,\cdots,-4}_{5},~0),\nn\\
\eeqa
with normalization factor $c=3$.

After the ${\bf (~45,~1)}$ component of ${\bf 78}$ dimensional Higgs acquire VEVs which is denoted by $\langle{\bf 78}\rangle_{\bf (45,1)}$, the new contributions to superpotential
\beqa W  &\supset &\f{v_{\bf 78}}{2\sqrt{10}M_*} \sum\limits_{i,j=1}^3\[ -h_{ij}^a{\bf 10}_i {\bf 10}_j{\bf 5}_H+h_{ij}^a{\bf 10}^i_m {\bf \bar{5}}^j_m{\bf \bar{5}}_H-2h_{ij}^a{\bf \bar{5}}_m{\bf 1}_m{\bf 5}_H\.\nn\\&&~~~~~+2 h_{ij}^{\pr s}{\bf 10}_i {\bf 10}_j{\bf 5}_H-4h_{ij}^{\pr s}{\bf 10}^i_m {\bf \bar{5}}^j_m{\bf \bar{5}}_H+4h_{ij}^{\pr s}{\bf \bar{5}}_m{\bf 1}_m{\bf 5}_H\left.\f{}{}\right],\nn\\
&\supset &\f{v_{\bf 78}}{2\sqrt{10}M_*} \sum\limits_{i,j=1}^3\[4h_{ij}^{\pr s}Q_L^i(D_L^c)^jH_d+(-2h_{ij}^a+4h_{ij}^{\pr s})L_L^i(N_L^c)^jH_u\.\nn\\&&~~~~~~~~~~~+(h_{ij}^a-4h_{ij}^{\pr s}) \{Q_L^i(U_L^c)^jH_u+ L_L^i(E_L^c)^jH_d\}\left.\f{}{}\right],
\eeqa
while the new contributions to supersymmetry breaking soft trilinear terms
 \beqa
 -{\cal L} &\supset& \al'\f{v_{\bf 78}F_T}{2\sqrt{10}M_*^2}\sum\limits_{i,j=1}^3\[ 4y_{ij}^{\pr s}\tl{Q}_L^i(\tl{D}_L^c)^jH_d+ (-2y_{ij}^a+4y_{ij}^{\pr s})\tl{L}_L^i(\tl{N}_L^c)^jH_u\}\.\nn\\&&~~~~~~~~+ (y_{ij}^a-4y_{ij}^{\pr s})\{\tl{Q}_L^i(\tl{U}_L^c)^jH_u+ \tl{L}_L^i(\tl{E}_L^c)^jH_d\}\left.\f{}{}\right]. \eeqa

After ${\bf (~45,~1)}$ component of ${\bf 650}$ dimensional Higgs acquire VEVs which is denoted by $\langle{\bf 650}\rangle_{\bf (45,1)}$, the new contributions to superpotential
\beqa W  &\supset &\f{v_{650}}{4\sqrt{5}M_*} \sum\limits_{i,j=1}^3\[ h_{ij}^a{\bf 10}_i {\bf 10}_j{\bf 5}_H+5h_{ij}^a{\bf 10}^i_m {\bf \bar{5}}^j_m{\bf \bar{5}}_H-4h_{ij}^a{\bf \bar{5}}_m{\bf 1}_m{\bf 5}_H\.\nn\\&&~~~~~~~+ h_{ij}^{\pr s}{\bf 10}_i {\bf 10}_j{\bf 5}_H-3h_{ij}^{\pr s}{\bf 10}^i_m {\bf \bar{5}}^j_m{\bf \bar{5}}_H-4h_{ij}^{\pr s}{\bf \bar{5}}_m{\bf 1}_m{\bf 5}_H\nn\\
&&~~~~~~~+4 h_{ij}^{\pr\pr s}{\bf 10}_i {\bf 10}_j{\bf 5}_H-8h_{ij}^{\pr\pr s}{\bf 10}^i_m {\bf \bar{5}}^j_m{\bf \bar{5}}_H+8h_{ij}^{\pr\pr s}{\bf \bar{5}}_m{\bf 1}_m{\bf 5}_H\left.\f{}{}\right],\nn\\
&\supset &\f{v_{650}}{4\sqrt{5}M_*} \sum\limits_{i,j}^3\[\f{}{}(h_{ij}^{\pr s}+4h_{ij}^{\pr\pr s})Q_L^i(D_L^c)^jH_d\.\nn\\&&~~~~+(-4h_{ij}^a-4h_{ij}^{\pr s}+8h_{ij}^{\pr\pr s})L_L^i(N_L^c)^jH_u\nn\\&&~~~~+(5h_{ij}^a-3h_{ij}^{\pr s}-8h_{ij}^{\pr\pr s}) \{Q_L^i(U_L^c)^jH_u+ L_L^i(E_L^c)^jH_d\}\left.\f{}{}\right],
\eeqa
while the new contributions to supersymmetry breaking soft trilinear terms
 \beqa\scriptsize
 -{\cal L} &\supset& \al^\pr\f{v_{650}F_T}{4\sqrt{5}M_*^2}\sum\limits_{i,j=1}^3\[ (y_{ij}^{\pr s}+4y_{ij}^{\pr\pr s})\tl{Q}_L^i(\tl{D}_L^c)^jH_d\.\nn\\&&~~~-(4y_{ij}^a+4y_{ij}^{\pr s}-8y_{ij}^{\pr\pr s})\tl{L}_L^i(\tl{N}_L^c)^jH_u\}\nn\\&&~~+ (5y_{ij}^a-3y_{ij}^{\pr s}-8y_{ij}^{\pr\pr s})\{\tl{Q}_L^i(\tl{U}_L^c)^jH_u+\tl{L}_L^i(\tl{E}_L^c)^jH_d\}\left.\f{}{}\right]. \eeqa\normalsize

\subsection{$E_6$ To $SU(3)_C\tm SU(3)_L\tm SU(3)_R$ Model}
  The gauge invariant Yukawa coupling in $E_6$ GUT have the form
  \beqa
  W&\supset& \sum\limits_{i,j=1}^3y_{ij} {\bf 27}^i{\bf 27}^j{\bf 27}_h\supset \sum\limits_{i,j}^3\(2y_{ij}^sX_L^i (X_L^c)^jH+y_{ij}N^i N^jH\)\nn~\\
   &\supset& \sum\limits_{i,j=1}^3y_{ij}^s\[\f{}{} 2Q_L^i(Q_L^c)^j\Phi+2L_L^i(L_L^c)^j\Phi\]\nn\\
   &\supset& \sum\limits_{i,j=1}^3y_{ij}^s \[2Q_L^i(U_L^c)^j h_u+ 2Q_L^i(D_L^c)^j h_d+ 2L_L^i(E_L^c)^j h_d+ 2L_L^i(N_L^c)^j h_u\].\nn\\
  \eeqa
 in which we identify the $({\bf 1_c,~\bar{3},~3})$ components of the Higgs fields ${\bf 27}_H$ as $H$; the bi-doublets $({\bf 1_c,~2,~2})$ in $H$ as $\Phi$; $y_{ij}$ is decomposed into symmetric $y_{ij}^s$ and antisymmetric $y_{ij}^a$ parts.

 We know that the breaking of $E_6$ into $SU(3)_C\tm SU(3)_L\tm SU(3)_R$ are realized by VEVs of ${\bf 650}$ dimensional representation Higgs fields.
As noted before, the left-right symmetric VEVs can be chosen as
\beqa
\small
<{\bf 650}>_1=\f{v_{\bf 650}}{3\sqrt{2}}{\rm diag}(\underbrace{-2,\cdots,-2}_{9},\underbrace{~1,\cdots,~1}_{9}, \underbrace{~1,\cdots,~1}_{9})~,
\eeqa
\normalsize
while the left-right non-symmetric VEVs can be chosen as
\beqa
<{\bf 650}>_2=\f{\tl{v}_{\bf 650}}{\sqrt{6}}{\rm diag}(\underbrace{~0,\cdots,~0}_{9},\underbrace{~1,\cdots,~1}_{9}, \underbrace{-1,\cdots,-1}_{9})~,
\eeqa
with normalization factor $c=3$.

 After ${\bf 650}$ dimensional Higgs acquire left-right symmetric VEVs $\langle{\bf 650}\rangle_1$, the new contributions to superpotential
\beqa\scriptsize W &\supset&  \f{v_{\bf 650}}{3\sqrt{2}M_*}\sum\limits_{i,j=1}^3\left\{\f{}{}\. h_{ij}^{\pr s}\[2X_L^i (X_L^c)^j H-2N^iN^jH\]\nn\\& &~~~~~~~~~~~~~-2h_{ij}^{\pr\pr s}\[2X_L^i(X_L^c)^jH+ N^i N^j H\]\left.\f{}{}\right\},\nn\\
 &\supset&  \f{v_{\bf 650}}{3\sqrt{2}M_*}\sum\limits_{i,j=1}^3\left\{\f{}{}\. h_{ij}^{\pr s}\[2Q_L^i (Q_L^c)^j \Phi-4L_L^i(L_L^c)^j\Phi\]\nn\\&&~~~~~~~~~~~~-4h_{ij}^{\pr\pr s}\[Q_L^i(Q_L^c)^j\Phi+ L_L^i(L_L^c)^j\Phi\]\left.\f{}{}\right\},\nn\\
 &\supset& \f{v_{\bf 650}}{3\sqrt{2}M_*}\sum\limits_{i,j=1}^3\left[\f{}{} (2h_{ij}^{\pr s}-4 h_{ij}^{\pr\pr s}) \left\{2Q_L^i(U_L^c)^jh_u+2Q_L^i(D_L^c)^jh_d\right\} \.\nn\\&&~~~~~~~~
-4 (h_{ij}^{\pr s}+h_{ij}^{\pr\pr s})\left\{L_L^i(E_L^c)^jh_d+L_L^i(N_L^c)^jh_u\right\}\left.\f{}{}\right],
 \eeqa\normalsize
 while the new contributions to supersymmetry breaking soft trilinear terms
 \beqa
 -{\cal L} &\supset& \al'\f{v_{\bf 650}F_T}{3\sqrt{2}M_*^2}\sum\limits_{i,j=1}^3\left\{\f{}{} (2y_{ij}^{\pr s}-4 y_{ij}^{\pr\pr s})\[\tl{Q}_L^i(\tl{U}_L^c)^j h_u+\tl{Q}_L^i(\tl{D}_L^c)^j h_d\]\.\nn\\
  & & ~~~~~~~~~~~~~~-4(y_{ij}^{\pr s}+y_{ij}^{\pr\pr s})\[\tl{L}_L^i(\tl{E}_L^c)^j h_d+\tl{L}_L^i(\tl{N}_L^c)^j h_u\]\left.\f{}{}\right\}.
  \eeqa
The ${\bf 650}$ dimensional Higgs can also acquire left-right non-symmetric VEVs $\langle{\bf 650}\rangle_2$, so the new contribution to superpotential
\beqa W &\supset&  \f{\tl{v}_{\bf 650}}{\sqrt{6}M_*}\sum\limits_{i,j=1}^3\left\{\f{}{} 2h_{ij}^aX_L^i (X_L^c)^j H\f{}{}\right\},\nn\\
 &\supset& \f{\tl{v}_{\bf 650}}{\sqrt{6}M_*}\sum\limits_{i,j=1}^3\left[\f{}{} h_{ij}^a \left\{2Q_L^i(U_L^c)^jh_u+2Q_L^i(D_L^c)^jh_d \right\}\f{}{}\right],
 \eeqa
 while the new contributions to supersymmetry breaking soft trilinear terms
 \beqa
 -{\cal L} &\supset& \al'\f{\tl{v}_{\bf 650}F_T}{\sqrt{6}M_*^2}\sum\limits_{i,j=1}^3\left\{\f{}{} 2y_{ij}^a\tl{Q}_L^i(\tl{U}_L^c)^j h_u+2y_{ij}^a\tl{Q}_L^i(\tl{D}_L^c)^j h_d \f{}{}\right\}.
  \eeqa

\subsection{$E_6$ To $SU(3)_C\tm SU(2)_L\tm SU(2)_R\tm U(1)_1\tm U(1)_2$ Model}
We know that this symmetry broken chain can be realized by the VEVs of ${\bf (~1, ~8,~8)}$ components in ${\bf 650}$ dimensional representation Higgs fields
\beqa
\small
\langle{\bf 650}\rangle=\f{\hat{v}_{\bf 650}}{2\sqrt{3}}{\rm diag}({~1,~1,-2,~1,~1,-2,-2,-2,~4}, \underbrace{~0,\cdots,~0}_{9},\underbrace{~0,\cdots,~0}_{9}).\nn\\
\eeqa\normalsize
with normalization $c=3$.
After ${\bf 650}$ dimensional Higgs acquire left-right symmetric VEVs $\langle{\bf 650}\rangle$, the new contributions to superpotential
\beqa W &\supset& \f{\hat{v}_{\bf 650}}{2\sqrt{3}M_*}\sum\limits_{i,j=1}^3\left\{-4h_{ij}^{\pr s}L_L^i(L_L^c)^j\Phi+h_{ij}^{\pr\pr s}\[2Q_L^i(Q_L^c)^j\Phi+2L_L^i(L_L^c)^j\Phi\]\f{}{}\right\},\nn\\
&\supset& \f{\hat{v}_{\bf 650}}{2\sqrt{3}M_*}\sum\limits_{i,j=1}^3\[\f{}{}(-4h_{ij}^{\pr s}+2h_{ij}^{\pr \pr s})\left\{L_L^i(E_L^c)^jh_d+L_L^i(N_L^c)^jh_u\right\} \. \nn\\
&&~~~~~~~~~~~~~~~~~~~~~+2h_{ij}^{\pr \pr s}\{Q_L^i(U_L^c)^jh_u+Q_L^i(D_L^c)^jh_d \}\left.\f{}{}\].
 \eeqa
 while the new contributions to supersymmetry breaking soft trilinear terms
 \beqa
 -{\cal L} &\supset& \al'\f{\hat{v}_{\bf 650}F_T}{2\sqrt{3}M_*^2}\sum\limits_{i,j=1}^3\left\{\f{}{}(-4y_{ij}^{\pr s}+2y_{ij}^{\pr\pr s})\[\tl{L}_L^i(\tl{E}_L^c)^j h_d
+\tl{L}_L^i(\tl{N}_L^c)^j h_u \]\.\nn\\
  & &~~~~~~~~~~~~~~~~~~~~~~~+2y_{ij}^{\pr\pr s}\[\tl{Q}_L^i(\tl{U}_L^c)^jh_u+\tl{Q}_L^i(\tl{D}_L^c)^jh_d \] \left.\f{}{}\right\}~.
\eeqa

As noted before, this symmetry broken chain can also be realized by the VEVs of both $({\bf ~1,~1,~8})$ and $({\bf ~1,~8,~1})$ components of ${\bf 78}$ dimensional representation Higgs fields
\beqa\small
\langle{\bf 78}\rangle_1&=&\f{{v}_{\bf 78}}{\sqrt{6}}{\rm diag}(\underbrace{-1,-1,~2}_{3},\underbrace{~1,~1,-2}_{3},\underbrace{~0,\cdots,~0}_{9})~,\\
\langle{\bf 78}\rangle_2&=&\f{\tl{v}_{\bf 78}}{\sqrt{6}}{\rm diag}(\underbrace{-1,\cdots,-1}_{6},\underbrace{~2,\cdots,~2}_{3},\underbrace{~0,\cdots,~0}_{9},\underbrace{~1,~1,-2}_{3}),
\eeqa\normalsize
with $c=3$.

Besides, it is also possible for this symmetry broken chain to be realized by the VEVs of both $({\bf ~1,~1,~8})$ and $({\bf ~1,~8,~1})$ components of ${\bf 650}$ dimensional representation Higgs fields
\beqa\small
\langle{\bf 650}\rangle_1&=&\f{\hat{v}_{\bf 650}^\pr}{\sqrt{6}}{\rm diag}(\underbrace{~1,~1,-2}_{3},\underbrace{~1,~1,-2}_{3},\underbrace{~0,\cdots,~0}_{9})~,\\
\langle{\bf 650}\rangle_2&=&\f{\hat{v}^{\pr\pr}_{\bf 650}}{\sqrt{6}}{\rm diag}(\underbrace{~1,\cdots,~1}_{6},\underbrace{-2,\cdots,-2}_{3},\underbrace{~0,\cdots,~0}_{9},\underbrace{~1,~1,-2}_{3})~,
\eeqa\normalsize
with $c=3$.

 After both $({\bf ~1,~1,~8})$ and $({\bf ~1,~8,~1})$ components in ${\bf 78}$ dimensional Higgs acquire VEVs, the new contributions to superpotential
\beqa W &\supset& \f{v_{\bf 78}}{\sqrt{6}M_*}\left\{\f{}{} h_{1ij}^a Q_L^i (Q_L^c)^j\Phi- 3h_{1ij}^a L_L^i (L_L^c)^j\Phi+  h_{1ij}^{\pr s} Q_L^i (Q_L^c)^j\Phi \.\nn\\
&&~~~~~~~~~+h_{1ij}^{\pr s} L_L^i (L_L^c)^j \Phi -2 h_{1ij}^{\pr\pr s} Q_L^i (Q_L^c)^j\Phi-2 h_{1ij}^{\pr\pr s} L_L^i (L_L^c)^j\Phi\left.\right\},\nn\\
&&+\f{\tl{v}_{\bf 78}}{\sqrt{6}M_*}\left\{\f{}{} -h_{2ij}^a Q_L^i (Q_L^c)^j\Phi + 3h_{2ij}^a L_L^i (L_L^c)^j\Phi+  h_{2ij}^{\pr s} Q_L^i (Q_L^c)^j\Phi \.\nn\\
&&~~~~~~~~~+h_{2ij}^{\pr s} L_L^i (L_L^c)^j \Phi -2 h_{2ij}^{\pr\pr s} Q_L^i (Q_L^c)^j\Phi-2 h_{2ij}^{\pr\pr s} L_L^i (L_L^c)^j\Phi\left.\right\},\nn\\
&\supset& \f{{v}_{\bf 78}}{\sqrt{6}M_*}\[\f{}{}(h_{1ij}^a+h_{1ij}^{\pr s}-2h_{1ij}^{\pr\pr s})\{Q_L^i(U_L^c)^jh_u+Q_L^i(D_L^c)^jh_d\}\.\nn\\&&~~~~~~+(-3h_{1ij}^a+h_{1ij}^{\pr s}-2h_{1ij}^{\pr\pr s})\left\{L_L^i(E_L^c)^jh_d+L_L^i(N_L^c)^jh_u \right\}\left.\f{}{}\]\nn\\
& &+ \f{\tl{v}_{\bf 78}}{\sqrt{6}M_*}\[\f{}{}(-h_{2ij}^a+h_{2ij}^{\pr s}-2h_{2ij}^{\pr\pr s})\{Q_L^i(U_L^c)^jh_u+Q_L^i(D_L^c)^jh_d\}\.\nn\\&&~~~~~~+(3h_{2ij}^a+h_{2ij}^{\pr s}-2h_{2ij}^{\pr\pr s})\left\{L_L^i(E_L^c)^jh_d+L_L^i(N_L^c)^jh_u \right\}\left.\f{}{}\].
\eeqa
 while the supersymmetry breaking soft trilinear terms
 \beqa
 -{\cal L} &\supset& \al'\f{{v}_{\bf 78}F_T}{\sqrt{6}M_*^2}\[\f{}{}(y_{1ij}^a+y_{1ij}^{\pr s}-2y_{1ij}^{\pr\pr s})\{\tl{Q}_L^i(\tl{U}_L^c)^jh_u+\tl{Q}_L^i(\tl{D}_L^c)^jh_d\}\.\nn\\&&~~~~~~+(-3y_{1ij}^a+y_{1ij}^{\pr s}-2y_{1ij}^{\pr\pr s})\left\{\tl{L}_L^i(\tl{E}_L^c)^jh_d+\tl{L}_L^i(\tl{N}_L^c)^jh_u \right\}\left.\f{}{}\]\nn\\
&&+\al'\f{\tl{v}_{\bf 78}F_T}{\sqrt{6}M_*^2}\[\f{}{}(-y_{2ij}^a+y_{2ij}^{\pr s}-2y_{2ij}^{\pr\pr s})\left\{\tl{Q}_L^i(\tl{U}_L^c)^jh_u+\tl{Q}_L^i(\tl{D}_L^c)^jh_d\right\}\.\nn\\&&~~~~~~+(3y_{2ij}^a+y_{2ij}^{\pr s}-2y_{2ij}^{\pr\pr s})\left\{\tl{L}_L^i(\tl{E}_L^c)^jh_d+\tl{L}_L^i(\tl{N}_L^c)^jh_u \right\}\left.\f{}{}\].
  \eeqa


Similarly, after both $({\bf ~1,~1,~8})$ and $({\bf ~1,~8,~1})$ components in ${\bf 650}$ dimensional Higgs acquire VEVs, the new contributions to superpotential

\beqa\small W &\supset& \f{v_{\bf 650}}{\sqrt{6}M_*}\left\{\f{}{} h_{1ij}^a Q_L^i (Q_L^c)^j\Phi+ 3h_{1ij}^a L_L^i (L_L^c)^j\Phi+  h_{1ij}^{\pr s} Q_L^i (Q_L^c)^j\Phi \.\nn\\
&&~~~~~~~~~-h_{1ij}^{\pr s} L_L^i (L_L^c)^j \Phi +2 h_{1ij}^{\pr\pr s} Q_L^i (Q_L^c)^j\Phi+2 h_{1ij}^{\pr\pr s} L_L^i (L_L^c)^j\Phi\left.\right\},\nn\\
&&+\f{\tl{v}_{\bf 650}}{\sqrt{6}M_*}\left\{\f{}{} -h_{2ij}^a Q_L^i (Q_L^c)^j\Phi - 3h_{2ij}^a L_L^i (L_L^c)^j\Phi+  h_{2ij}^{\pr s} Q_L^i (Q_L^c)^j\Phi \.\nn\\
&&~~~~~~~~~-h_{2ij}^{\pr s} L_L^i (L_L^c)^j \Phi +2 h_{2ij}^{\pr\pr s} Q_L^i (Q_L^c)^j\Phi+2 h_{2ij}^{\pr\pr s} L_L^i (L_L^c)^j\Phi\left.\right\},\nn\\
&\supset& \f{{v}_{\bf 650}}{\sqrt{6}M_*}\[\f{}{}(h_{1ij}^a+h_{1ij}^{\pr s}+2h_{1ij}^{\pr\pr s})\{Q_L^i(U_L^c)^jh_u+Q_L^i(D_L^c)^jh_d\}\.\nn\\&&~~~~~~+(3h_{1ij}^a-h_{1ij}^{\pr s}+2h_{1ij}^{\pr\pr s})\left\{L_L^i(E_L^c)^jh_d+L_L^i(N_L^c)^jh_u \right\}\left.\f{}{}\]\nn\\
& &+ \f{\tl{v}_{\bf 650}}{\sqrt{6}M_*}\[\f{}{}(-h_{2ij}^a+h_{2ij}^{\pr s}+2h_{2ij}^{\pr\pr s})\{Q_L^i(U_L^c)^jh_u+Q_L^i(D_L^c)^jh_d\}\.\nn\\&&~~~~~~+(-3h_{2ij}^a-h_{2ij}^{\pr s}+2h_{2ij}^{\pr\pr s})\left\{L_L^i(E_L^c)^jh_d+L_L^i(N_L^c)^jh_u \right\}\left.\f{}{}\],
\eeqa\normalsize
  while the new contributions to supersymmetry breaking soft trilinear terms
 \beqa
 -{\cal L} &\supset& \al'\f{{v}_{\bf 78}F_T}{\sqrt{6}M_*^2}\[(y_{1ij}^a+y_{1ij}^{\pr s}+2y_{1ij}^{\pr\pr s})\{\tl{Q}_L^i(\tl{U}_L^c)^jh_u+\tl{Q}_L^i(\tl{D}_L^c)^jh_d\}\.\nn\\&&~~~~+(3y_{1ij}^a-y_{1ij}^{\pr s}+2y_{1ij}^{\pr\pr s})\left\{\tl{L}_L^i(\tl{E}_L^c)^jh_d+\tl{L}_L^i(\tl{N}_L^c)^jh_u \right\}\left.\f{}{}\]\nn\\
& &+\al' \f{\tl{v}_{\bf 78}F_T}{\sqrt{6}M_*^2}\[(-y_{2ij}^a+y_{2ij}^{\pr s}+2y_{2ij}^{\pr\pr s})\{\tl{Q}_L^i(\tl{U}_L^c)^jh_u+\tl{Q}_L^i(\tl{D}_L^c)^jh_d\}\.\nn\\&&~~+(-3y_{2ij}^a-y_{2ij}^{\pr s}+2y_{2ij}^{\pr\pr s})\left\{\tl{L}_L^i(\tl{E}_L^c)^jh_d+\tl{L}_L^i(\tl{N}_L^c)^jh_u \right\}\left.\f{}{}\].
  \eeqa

\subsection{$E_6$ To $SU(6)\tm SU(2)$ Model}
We know that this symmetry broken chain can be realized via the VEVs of ${\bf 650}$ dimensional representation Higgs field.
The VEVs that break $E_6$ into $SU(6)\tm SU(2)$ can be chosen as
\beqa
<{\bf 650}>=\f{v_{\bf 650}}{6\sqrt{5}}{\rm diag}(\underbrace{-4,\cdots,-4}_{15},\underbrace{~5,\cdots,~5}_{12})~,
\eeqa
with normalization factor $c=3$.
\begin{itemize}
\item $E_6\ra SU(6)\tm SU(2)_X \ra SU(5) \tm U(1) \tm SU(2)_X$:
The filling of matter contents can be seen in previous sections.
 The gauge invariant (renormalizable) Yukawa coupling in $E_6$ GUT thus have the form
  \beqa
  W&\supset& \sum\limits_{i,j=1}^3y_{ij} {\bf 27}^i{\bf 27}^j{\bf 27}_h ~,\nn~\\
   &\supset & \sum\limits_{i,j=1}^3\(y_{ij}^s F_{\bf (10,1)}^i F_{\bf (10,1)}^j H_{\bf (5,1)}+2y_{ij}^s F_{\bf (10,1)}^i F_{\bf (\bar{5},2)}^j H_{\bf(\bar{5},2)}\.\nn\\&&~~~~~~~+ 2y_{ij}^s F_{\bf (\bar{5},2)}^i F_{\bf (1,2)}^j H_{\bf(5,1)}\left.\f{}{}\)~,\nn\\
   &\supset& \sum\limits_{i,j=1}^3y_{ij}^s\[\f{}{} 2Q_L^i(U_L^c)^j h_u+ 2Q_L^i(D_L^c)^j h_d+ 2L_L^i(E_L^c)^j h_d\.\nn\\&&~~~~~~~~~+ 2L_L^i(N_L^c)^j h_u\left.\f{}{}\].
   \eeqa

After ${\bf 650}$ dimensional Higgs acquire VEVs $\langle{\bf 650}\rangle$, the new contributions to superpotential
\beqa W  &\supset &\f{v_{\bf 650}}{6\sqrt{5}M_*} \sum\limits_{i,j=1}^3\(-9h_{ij}^a F_{\bf (10,1)}^i F_{\bf (\bar{5},2)}^j H_{\bf(\bar{5},2)}\f{}{}\)~,\nn\\
 &+&\f{v_{\bf 650}}{6\sqrt{5}M_*} \sum\limits_{i,j=1}^3\(-4h_{ij}^{\pr s} F_{\bf (10,1)}^i F_{\bf (10,1)}^j H_{\bf (5,1)}+h_{ij}^{\pr s} F_{\bf (10,1)}^i F_{\bf (\bar{5},2)}^j H_{\bf(\bar{5},2)}\.\nn\\& &~~~~~~~~+10h_{ij}^{\pr\pr s} F_{\bf (\bar{5},2)}^i F_{\bf (1,2)}^j H_{\bf(5,1)}\left.\f{}{}\)~,\nn\\
 &+&\f{v_{\bf 650}}{6\sqrt{5}M_*} \sum\limits_{i,j=1}^3\(-4h_{ij}^{\pr\pr s} F_{\bf (10,1)}^i F_{\bf (10,1)}^j H_{\bf (5,1)}+10h_{ij}^{\pr\pr s} F_{\bf (10,1)}^i F_{\bf (\bar{5},2)}^j H_{\bf(\bar{5},2)}\.\nn\\& &~~~~~~~~
 -8h_{ij}^{\pr\pr s} F_{\bf (\bar{5},2)}^i F_{\bf (1,2)}^j H_{\bf(5,1)}\left.\f{}{}\)~,\nn\\
 &\supset& \f{v_{\bf 650}}{6\sqrt{5}M_*} \sum\limits_{i,j=1}^3\[\f{}{} -8(h_{ij}^{\pr s}+h_{ij}^{\pr\pr s}) Q_L^i(U_L^c)^j h_u\.\nn\\&&~~~~~~~~~~~~~~+(-9h_{ij}^a+h_{ij}^{\pr s}+10h_{ij}^{\pr\pr s})Q_L^i(D_L^c)^j h_d\left.\f{}{}\],\nn\\
 &+&\f{v_{\bf 650}}{6\sqrt{5}M_*} \sum\limits_{i,j=1}^3\[\f{}{}(10h_{ij}^{\pr s}-8h_{ij}^{\pr\pr s})L_L^i(N_L^c)^j h_u\.\nn\\&&~~~~~~~~~~~~~~+   (-9h_{ij}^a+h_{ij}^{\pr s}+10h_{ij}^{\pr\pr s})L_L^i(E_L^c)^j h_d\left.\f{}{}\],
   \eeqa
 while the new contributions to supersymmetry breaking soft trilinear terms
 \beqa
 -{\cal L} &\supset& \al'\f{v_{\bf 650}F_T}{6\sqrt{5}M_*^2} \sum\limits_{i,j=1}^3\[\f{}{}-8(y_{ij}^{\pr s}+y_{ij}^{\pr\pr s})\tl{Q}_L^i(\tl{U}_L^c)^j h_u\.\nn\\&&~~~~~~~~~~~~~~+ (-9y_{ij}^a+y_{ij}^{\pr s}+10y_{ij}^{\pr\pr s})\tl{Q}_L^i(\tl{D}_L^c)^j h_d\left.\f{}{}\],\nn\\
 &+&\al'\f{v_{\bf 650}F_T}{6\sqrt{5}M_*^2} \sum\limits_{i,j=1}^3\[\f{}{}(10y_{ij}^{\pr s}-8y_{ij}^{\pr\pr s})\tl{L}_L^i(\tl{N}_L^c)^j h_u \.\nn\\&&~~~~~~~~~~~~~~+(-9y_{ij}^a+y_{ij}^{\pr s}+10y_{ij}^{\pr\pr s})\tl{L}_L^i(\tl{E}_L^c)^j h_d\left.\f{}{}\].
  \eeqa

\item $E_6\ra SU(6)\tm SU(2)_L\ra SU(4)_c\tm SU(2)_L\tm SU(2)_R \tm U(1)_1$:
 The gauge invariant Yukawa coupling in $E_6$ GUT thus have the form
  \beqa
  W&\supset& \sum\limits_{i,j=1}^3y_{ij} {\bf 27}^i{\bf 27}^j{\bf 27}_h ~,\nn~\\
   &\supset & \sum\limits_{i,j=1}^3\(2y_{ij}^s F_{\bf (4,2,1)}^i F_{\bf (\bar{4},1,2)}^j H_{\bf (1,\bar{2},2)}\)~,\nn\\
   &\supset& \sum\limits_{i,j=1}^3y_{ij}^s\[\f{}{} 2Q_R^i(Q_R^c)^j \Phi + 2L_R^i(L_R^c)^j \Phi \]~,\nn\\
   &\supset& -\sum\limits_{i,j=1}^3y_{ij}^s\{2Q_L^i(U_L^c)^jH_u+2Q_L^i(D_L^c)^jH_d\nn\\&&~~~~~~ + 2L_L^i(E_L^c)^jH_d+ 2L_L^i(N_L^c)^jH_u\}.
  \eeqa

After ${\bf 650}$ dimensional Higgs acquire VEVs $\langle{\bf 650}\rangle_1$, the new contributions to superpotential
\beqa W  &\supset &\f{v_{\bf 650}}{6\sqrt{5}M_*} \sum\limits_{i,j}^3\[ -9h_{ij}^a F_{\bf (4,2,1)}^i F_{\bf (\bar{4},1,2)}^j H_{\bf (1,\bar{2},2)}\.\nn\\&&~
+h_{ij}^{\pr s}F_{\bf (4,2,1)}^i F_{\bf (\bar{4},1,2)}^j H_{\bf (1,\bar{2},2)}+10h_{ij}^{\pr\pr s} F_{\bf (4,2,1)}^i F_{\bf (\bar{4},1,2)}^j H_{\bf (1,\bar{2},2)}\left.\f{}{}\right],\nn\\
&\supset &\f{v_{\bf 650}}{6\sqrt{5}M_*} \sum\limits_{i,j}^3\[\f{}{}(9h_{ij}^a+h_{ij}^{\pr s}+10h_{ij}^{\pr\pr s}) \left\{\f{}{}2Q_L^i(U_L^c)^jH_u\.\.\nn\\&&~~~+2Q_L^i(D_L^c)^jH_d + 2L_L^i(E_L^c)^jH_d+ 2L_L^i(N_L^c)^jH_u\left.\f{}{}\right\} \left.\f{}{}\right],\eeqa
while the new contributions to supersymmetry breaking soft trilinear terms
 \beqa
 -{\cal L} &\supset& \al'\f{v_{\bf 650}F_T}{6\sqrt{5}M_*^2}\sum\limits_{i,j}^3\[ (9y_{ij}^a+ y_{ij}^{\pr s}+10y_{ij}^{\pr\pr s}) \left\{\f{}{}2\tl{Q}_L^i(\tl{U}_L^c)^jH_u\.\.\nn\\&&+2\tl{Q}_L^i(\tl{D}_L^c)^jH_d + 2\tl{L}_L^i(\tl{E}_L^c)^jH_d+ 2\tl{L}_L^i(\tl{N}_L^c)^jH_u\left.\f{}{}\right\}\left.\f{}{}\right]. \eeqa

\item $E_6\ra SU(6)\tm SU(2)_R\ra SU(4)_c\tm SU(2)_L\tm SU(2)_R \tm U(1)_1$:

The gauge invariant Yukawa coupling in $E_6$ GUT thus have the form
  \beqa
  W&\supset& \sum\limits_{i,j=1}^3y_{ij} {\bf 27}^i{\bf 27}^j{\bf 27}_h ~,\nn~\\
   &\supset & \sum\limits_{i,j=1}^3\(2y_{ij}^s F_{\bf (4,2,1)}^i F_{\bf (\bar{4},1,2)}^j H_{\bf (1,\bar{2},2)}\)~,\nn\\
   &\supset& \sum\limits_{i,j=1}^3y_{ij}^s\[\f{}{} 2Q_L^i(Q_L^c)^j \Phi + 2L_L^i(L_L^c)^j \Phi \]~,\nn\\
   &\supset& \sum\limits_{i,j=1}^3y_{ij}^s\{2Q_L^i(U_L^c)^jH_u+2Q_L^i(D_L^c)^jH_d\nn\\&&~~~~~~ + 2L_L^i(E_L^c)^jH_d+ 2L_L^i(N_L^c)^jH_u\}~.
  \eeqa

After ${\bf 650}$ dimensional Higgs acquire VEVs $\langle{\bf 650}\rangle$, the new contributions to superpotential
\beqa W  &\supset &\f{v_{\bf 650}}{6\sqrt{5}M_*} \sum\limits_{i,j=1}^3\[ -9h_{ij}^a F_{\bf (4,2,1)}^i F_{\bf (\bar{4},1,2)}^j H_{\bf (1,\bar{2},2)}\.\nn\\&&~
+h_{ij}^{\pr s}F_{\bf (4,2,1)}^i F_{\bf (\bar{4},1,2)}^j H_{\bf (1,\bar{2},2)}+10h_{ij}^{\pr\pr s} F_{\bf (4,2,1)}^i F_{\bf (\bar{4},1,2)}^j H_{\bf (1,\bar{2},2)}\left.\f{}{}\right],\nn\\
&\supset &\f{v_{\bf 650}}{6\sqrt{5}M_*} \sum\limits_{i,j=1}^3\[(-9h_{ij}^a+h_{ij}^{\pr s}+10h_{ij}^{\pr\pr s})\left\{\f{}{}2Q_L^i(U_L^c)^jH_u\.\.\nn\\&&~~~~+2Q_L^i(D_L^c)^jH_d + 2L_L^i(E_L^c)^jH_d+ 2L_L^i(N_L^c)^jH_u\left.\f{}{}\right\} \left.\f{}{}\right].\eeqa
while the supersymmetry breaking soft trilinear terms
 \beqa
 -{\cal L} &\supset& \al'\f{v_{\bf 650}F_T}{6\sqrt{5}M_*^2}\sum\limits_{i,j=1}^3\[(-9y_{ij}^a+ y_{ij}^{\pr s}+10y_{ij}^{\pr\pr s})\left\{\f{}{}2\tl{Q}_L^i(\tl{U}_L^c)^jH_u\.\.\nn\\&&+2\tl{Q}_L^i(\tl{D}_L^c)^jH_d + 2\tl{L}_L^i(\tl{E}_L^c)^jH_d+ 2\tl{L}_L^i(\tl{N}_L^c)^jH_u\left.\f{}{}\right\}\left.\f{}{}\right]. \eeqa
\end{itemize}
\subsection{$E_6$ To $SU(4)_{c}\tm SU(2)_L\tm SU(2)_R\tm U(1)$ Model}
We know that this symmetry broken chain can be realized via the VEVs of ${\bf 650}$ as well as ${\bf 78}$ dimensional representation Higgs field.
The ${\bf (~35,~1)}$ component VEVs of the ${\bf 78}$ dimensional representation that break gauge group $SU(6)\tm SU(2)_1$ to its subgroup $SU(4)\tm SU(2)_1\tm SU(2)_2 \tm U(1)$ reads
\beqa\small
\langle{\bf 78}\rangle_{\bf (35,1)}=\f{v_{\bf 78}}{2\sqrt{6}}{\rm diag}(\underbrace{~1,~1,~1,~1,-2,-2}_{2},\underbrace{~2,~\cdots,~2}_{6},\underbrace{-1,~\cdots,-1}_{8},-4),\nn\\
\eeqa\normalsize
with normalization factor $c=3$.
The breaking of gauge group $SU(6)\tm SU(2)_1$ to its subgroup $SU(4)\tm SU(2)_1\tm SU(2)_2 \tm U(1)$ can be realized by both the ${\bf (35,1)}$ and the ${\bf (189,1)}$ component VEVs of ${\bf 650}$ dimensional representation
\beqa\small
\langle{\bf 650}\rangle_{\bf (35,1)}&=&\f{{v'}_{\bf 650}}{2\sqrt{3}}{\rm diag}(\underbrace{~1,~1,~1,~1,-2,-2}_{2},\underbrace{-1,~\cdots,-1}_{6},\underbrace{~\f{1}{2},~\cdots,~\f{1}{2}}_{8},~2),\nn\\
\langle{\bf 650}\rangle_{\bf (189,1)}&=&\f{\tl{v'}_{\bf 650}}{4\sqrt{5}}{\rm diag}(\underbrace{~0,\cdots,~0}_{12},\underbrace{-2,~\cdots,-2}_{6},\underbrace{~3,~\cdots,~3}_{8},-12),
\eeqa\normalsize
with normalization factor $c=3$.
\begin{itemize}
\item $E_6\ra SU(6)\tm SU(2)_L\ra SU(4)_c\tm SU(2)_L\tm SU(2)_R \tm U(1)_1$:

 After the ${\bf (~35,~1)}$ component of ${\bf 78}$ dimensional Higgs acquire VEVs $\langle{\bf 78}\rangle_{\bf (35,1)}$, the new contributions to superpotential
\beqa W  &\supset &\f{v_{\bf 78}}{2\sqrt{6}M_*} \sum\limits_{i,j=1}^3\[
-4h_{ij}^{\pr s} F_{\bf (4,2,1)}^i F_{\bf (\bar{4},1,2)}^j H_{\bf (1,\bar{2},2)}\f{}{}\right],\nn\\
&\supset &\f{v_{78}}{2\sqrt{6}M_*} \sum\limits_{i,j=1}^3\[\f{}{}-4h_{ij}^{\pr s} \{2Q_L^i(U_L^c)^jH_u\.\nn\\&&+2Q_L^i(D_L^c)^jH_d + 2L_L^i(E_L^c)^jH_d+ 2L_L^i(N_L^c)^jH_u\}\left.\f{}{}\right],
\eeqa
while the new contributions to supersymmetry breaking soft trilinear terms
 \beqa
 -{\cal L} &\supset& \al'\f{v_{\bf 78}F_T}{2\sqrt{6}M_*^2}\sum\limits_{i,j=1}^3\[ -4y_{ij}^{\pr s} \{2\tl{Q}_L^i(\tl{U}_L^c)^jH_u\.\nn\\&&+2\tl{Q}_L^i(\tl{D}_L^c)^jH_d + 2\tl{L}_L^i(\tl{E}_L^c)^jH_d+ 2\tl{L}_L^i(\tl{N}_L^c)^jH_u\}\left.\f{}{}\f{}{}\right]. \eeqa

After ${\bf (35,~1)}$ component of ${\bf 650}$ dimensional Higgs acquire VEVs which is denoted by $\langle{\bf 650}\rangle_{\bf (35,1)}$, the new contributions to superpotential
\beqa W  &\supset &\f{v^\pr_{\bf 650}}{2\sqrt{3}M_*} \sum\limits_{i,j=1}^3\[ -\f{3}{2}h_{ij}^aF_{\bf (4,2,1)}^i F_{\bf (\bar{4},1,2)}^j H_{\bf (1,\bar{2},2)}
\.\nn\\&&+\f{1}{2}h_{ij}^{\pr s} F_{\bf (4,2,1)}^i F_{\bf (\bar{4},1,2)}^j H_{\bf (1,\bar{2},2)}-4h_{ij}^{\pr\pr s} F_{\bf (4,2,1)}^i F_{\bf (\bar{4},1,2)}^j H_{\bf (1,\bar{2},2)}\left.\f{}{}\right],\nn\\
&\supset &\f{v_{\bf 650}^\pr}{2\sqrt{3}M_*} \sum\limits_{i,j=1}^3\[ (\f{3}{2}h_{ij}^a+\f{1}{2}h_{ij}^{\pr s}-4h_{ij}^{\pr\pr s})\left\{\f{}{}2Q_L^i(U_L^c)^jH_u\.\.\nn\\&&~~+2Q_L^i(D_L^c)^jH_d + 2L_L^i(E_L^c)^jH_d+ 2L_L^i(N_L^c)^jH_u\left.\f{}{}\right\}\left.\f{}{}\right],
\eeqa
while the new contributions to supersymmetry breaking soft trilinear terms
 \beqa
 -{\cal L} &\supset& \al'\f{v_{\bf 650}^\pr F_T}{2\sqrt{3}M_*^2}\sum\limits_{i,j=1}^3\[ (\f{3}{2}y_{ij}^a +\f{1}{2}y_{ij}^{\pr s}-4y_{ij}^{\pr\pr s})\left\{\f{}{}2\tl{Q}_L^i(\tl{U}_L^c)^jH_u\.\.\nn\\&&+2\tl{Q}_L^i(\tl{D}_L^c)^jH_d+ 2\tl{L}_L^i(\tl{E}_L^c)^jH_d+ 2\tl{L}_L^i(\tl{N}_L^c)^jH_u\left.\f{}{}\right\}\left.\f{}{}\right]. \eeqa

After ${\bf (189,1)}$ component of ${\bf 650}$ dimensional Higgs acquire VEVs which is denoted by $\langle{\bf 650}\rangle_{\bf (189,1)}$, the new contributions to superpotential
\beqa W  &\supset &\f{\tl{v'}_{\bf 650}}{6\sqrt{5}M_*} \sum\limits_{i,j=1}^3\[\f{}{} -3h_{ij}^a F_{\bf (4,2,1)}^i F_{\bf (\bar{4},1,2)}^j H_{\bf (1,\bar{2},2)}\.\nn\\&&~~~~~~~~~~~~~
-3h_{ij}^{\pr s} F_{\bf (4,2,1)}^i F_{\bf (\bar{4},1,2)}^j H_{\bf (1,\bar{2},2)}\left.\f{}{}\right],\nn\\
&\supset &\f{\tl{v'}_{\bf 650}}{6\sqrt{5}M_*} \sum\limits_{i,j=1}^3\[(3h_{ij}^a-3h_{ij}^{\pr s})\left\{\f{}{}2Q_L^i(U_L^c)^jH_u\.\.\nn\\&&+2Q_L^i(D_L^c)^jH_d + 2L_L^i(E_L^c)^jH_d+ 2L_L^i(N_L^c)^jH_u\left.\f{}{}\right\}
\left.\f{}{}\right],
\eeqa
while the new contributions to supersymmetry breaking soft trilinear terms
 \beqa
 -{\cal L} &\supset& \al'\f{\tl{v'}_{\bf 650}F_T}{6\sqrt{5}M_*^2}\sum\limits_{i,j=1}^3\[(3y_{ij}^a- 3y_{ij}^{\pr s})\left\{\f{}{}2\tl{Q}_L^i(\tl{U}_L^c)^jH_u\.\.\nn\\&&+2\tl{Q}_L^i(\tl{D}_L^c)^jH_d + 2\tl{L}_L^i(\tl{E}_L^c)^jH_d+ 2\tl{L}_L^i(\tl{N}_L^c)^jH_u\left.\f{}{}\right\}\left.\f{}{}\right].\eeqa

\item $E_6\ra SU(6)\tm SU(2)_R\ra SU(4)_c\tm SU(2)_L\tm SU(2)_R \tm U(1)_1$:

 After the ${\bf (~35,~1)}$ component of ${\bf 78}$ dimensional Higgs acquire VEVs which is denoted by $\langle{\bf 78}\rangle_{\bf (35,1)}$, the new contributions to superpotential
\beqa W  &\supset &\f{v_{\bf 78}}{2\sqrt{6}M_*} \sum\limits_{i,j=1}^3\[
-4h_{ij}^{\pr s} F_{\bf (4,2,1)}^i F_{\bf (\bar{4},1,2)}^j H_{\bf (1,\bar{2},2)}\f{}{}\right],\nn\\
&\supset &\f{v_{\bf 78}}{2\sqrt{6}M_*} \sum\limits_{i,j=1}^3\[-4h_{ij}^{\pr s} \{2Q_L^i(U_L^c)^jH_u\.\nn\\&&+2Q_L^i(D_L^c)^jH_d + 2L_L^i(E_L^c)^jH_d+ 2L_L^i(N_L^c)^jH_u\}\left.\f{}{}\right],
\eeqa
while the new contributions to supersymmetry breaking soft trilinear terms
 \beqa
 -{\cal L} &\supset& \al\f{v_{\bf 78}F_T}{2\sqrt{6}M_*^2}\sum\limits_{i,j=1}^3\[ -4y_{ij}^{\pr s} \{2\tl{Q}_L^i(\tl{U}_L^c)^jH_u\.\nn\\&&
 +2\tl{Q}_L^i(\tl{D}_L^c)^jH_d + 2\tl{L}_L^i(\tl{E}_L^c)^jH_d+ 2\tl{L}_L^i(\tl{N}_L^c)^jH_u\}\left.\f{}{}\right].\eeqa

After ${\bf (~35,~1)}$ component of ${\bf 650}$ dimensional Higgs acquire VEVs which is denoted by $\langle{\bf 650}\rangle_{\bf (35,1)}$, the new contributions to superpotential
\beqa W  &\supset &\f{v^\pr_{\bf 650}}{2\sqrt{3}M_*} \sum\limits_{i,j=1}^3\[ -\f{3}{2}h_{ij}^aF_{\bf (4,2,1)}^i F_{\bf (\bar{4},1,2)}^j H_{\bf (1,\bar{2},2)}
\.\nn\\&&~~+\f{1}{2}h_{ij}^{\pr s} F_{\bf (4,2,1)}^i F_{\bf (\bar{4},1,2)}^j H_{\bf (1,\bar{2},2)}-4h_{ij}^{\pr\pr s} F_{\bf (4,2,1)}^i F_{\bf (\bar{4},1,2)}^j H_{\bf (1,\bar{2},2)}\left.\f{}{}\right],\nn\\
&\supset &\f{v^\pr_{\bf 650}}{2\sqrt{3}M_*} \sum\limits_{i,j=1}^3\[(-\f{3}{2}h_{ij}^a +\f{1}{2}h_{ij}^{\pr s}-4h_{ij}^{\pr\pr s})\left\{\f{}{}2Q_L^i(U_L^c)^jH_u\.\.\nn\\&&~~~+2Q_L^i(D_L^c)^jH_d + 2L_L^i(E_L^c)^jH_d+ 2L_L^i(N_L^c)^jH_u\left.\f{}{}\right\}\left.\f{}{}\right],
\eeqa
while the new contributions to supersymmetry breaking soft trilinear terms
 \beqa
 -{\cal L} &\supset& \al'\f{v^\pr_{\bf 650}F_T}{2\sqrt{3}M_*^2}\sum\limits_{i,j=1}^3\[(-\f{3}{2}y_{ij}^a+\f{1}{2}y_{ij}^{\pr s}-4y_{ij}^{\pr\pr s}) \left\{\f{}{}2\tl{Q}_L^i(\tl{U}_L^c)^jH_u\.\.\nn\\&&+2\tl{Q}_L^i(\tl{D}_L^c)^jH_d + 2\tl{L}_L^i(\tl{E}_L^c)^jH_d+ 2\tl{L}_L^i(\tl{N}_L^c)^jH_u\left.\f{}{}\right\}\left.\f{}{}\right]. \eeqa

After ${\bf (189,1)}$ component of ${\bf 650}$ dimensional Higgs acquire VEVs which is denoted by $\langle{\bf 650}\rangle_{\bf (189,1)}$, the new contributions to superpotential
\beqa W  &\supset &\f{\tl{v'}_{\bf 650}}{6\sqrt{5}M_*} \sum\limits_{i,j=1}^3\[ -3h_{ij}^a F_{\bf (4,2,1)}^i F_{\bf (\bar{4},1,2)}^j H_{\bf (1,\bar{2},2)}
\.\nn\\&&~~~~~~~~~~~~~-3h_{ij}^{\pr s} F_{\bf (4,2,1)}^i F_{\bf (\bar{4},1,2)}^j H_{\bf (1,\bar{2},2)}\left.\f{}{}\right],\nn\\
&\supset &\f{\tl{v'}_{\bf 650}}{6\sqrt{5}M_*} \sum\limits_{i,j=1}^3\[(-3h_{ij}^a-3h_{ij}^{\pr s})\left\{\f{}{}2Q_L^i(U_L^c)^jH_u\.\.\nn\\&&+2Q_L^i(D_L^c)^jH_d + 2L_L^i(E_L^c)^jH_d+ 2L_L^i(N_L^c)^jH_u\left.\f{}{}\right\}
\left.\f{}{}\right],
\eeqa
while the new contributions to supersymmetry breaking soft trilinear terms
 \beqa
 -{\cal L} &\supset& \al'\f{\tl{v'}_{\bf 650}F_T}{6\sqrt{5}M_*^2}\sum\limits_{i,j=1}^3\[(-3y_{ij}^a- 3y_{ij}^{\pr s})\left\{\f{}{}2\tl{Q}_L^i(\tl{U}_L^c)^jH_u\.\.\nn\\&&+2\tl{Q}_L^i(\tl{D}_L^c)^jH_d + 2\tl{L}_L^i(\tl{E}_L^c)^jH_d+ 2\tl{L}_L^i(\tl{N}_L^c)^jH_u\left.\f{}{}\right\}\left.\f{}{}\right].\eeqa

\end{itemize}

\subsection{$E_6$ To $SU(5)\tm U(1)\tm SU(2)_X$}

The breaking of $E_6$ into $SU(5)\tm U(1) \tm SU(2)_X$ can be realized via the VEVs of ${\bf 78}$ and ${\bf 650}$ dimensional representations.
The ${\bf (~35,~1)}$ component VEVs of the ${\bf 78}$ dimensional representation that break $SU(6)\tm SU(2)_X$ gauge group to $SU(5)\tm U(1) \tm SU(2)_X$ reads
\beqa\small
\langle{\bf 78}\rangle_{\bf (35,1)}=\f{\hat{v}_{\bf 78}}{2\sqrt{15}}{\rm diag}(\underbrace{~1,~1,~1,~1,~1,-5}_{2},\underbrace{~2,~\cdots,~2}_{10},\underbrace{-4,~\cdots,-4}_{5}),
\eeqa\normalsize
with normalization factor $c=3$.
The ${\bf (~35,~1)}$ component VEVs of the ${\bf 650}$ dimensional representation that break $SU(6)\tm SU(2)_X$ gauge group to $SU(5)\tm U(1) \tm SU(2)_X$ reads
\beqa\small
\langle{\bf 650}\rangle_{\bf (35,1)}=\f{\hat{v}_{\bf 650}}{\sqrt{30}}{\rm diag}(\underbrace{~1,~1,~1,~1,~1,-5}_{2},\underbrace{-1,~\cdots,-1}_{10},\underbrace{~2,~\cdots,~2}_{5})~,
\eeqa\normalsize
with normalization factor $c=3$.

After ${\bf 78}$ dimensional Higgs acquire VEVs $\langle{\bf 78}\rangle_{\bf (35,1)}$, the new contributions to superpotential
\beqa W  &\supset &\f{\hat{v}_{\bf 78}}{2\sqrt{15}M_*} \sum\limits_{i,j=1}^3\(-3h_{ij}^a F_{\bf (10,1)}^i F_{\bf (\bar{5},2)}^j H_{\bf(\bar{5},2)}+6h_{ij}^a F_{\bf (\bar{5},2)}^i F_{\bf (1,2)}^j H_{\bf(5,1)}\f{}{}\)~,\nn\\
 &+&\f{\hat{v}_{\bf 78}}{2\sqrt{15}M_*} \sum\limits_{i,j=1}^3\(4h_{ij}^{\pr s} F_{\bf (10,1)}^i F_{\bf (10,1)}^j H_{\bf (5,1)}-10h_{ij}^{\pr s} F_{\bf (10,1)}^i F_{\bf (\bar{5},2)}^j H_{\bf(\bar{5},2)}\.\nn\\& &~~~~~~~~
+8h_{ij}^{\pr s} F_{\bf (\bar{5},2)}^i F_{\bf (1,2)}^j H_{\bf(5,1)}\left.\)~,\nn\\
 &\supset& \f{\hat{v}_{\bf 78}}{2\sqrt{15}M_*} \sum\limits_{i,j=1}^3\[\f{}{}(-3h_{ij}^a-10h_{ij}^{\pr s})\{Q_L^i(D_L^c)^j h_d+L_L^i(E_L^c)^j h_d\}\]\nn~,\\
 &+&\f{\hat{v}_{\bf 78}}{2\sqrt{15}M_*} \sum\limits_{i,j=1}^3\[\f{}{} 8h_{ij}^{\pr s} Q_L^i(U_L^c)^j h_u+(6h_{ij}^a+8h_{ij}^{\pr s})L_L^i(N_L^c)^j h_u\f{}{}\],
   \eeqa
 while the new contributions to supersymmetry breaking soft trilinear terms
 \beqa
 -{\cal L} &\supset& \al'\f{\hat{v}_{\bf 78}F_T}{2\sqrt{15}M_*^2} \sum\limits_{i,j=1}^3\[\f{}{}(-3y_{ij}^a-10y_{ij}^{\pr s})\{\tl{Q}_L^i(\tl{D}_L^c)^j h_d
+\tl{L}_L^i(\tl{E}_L^c)^j h_d\}\]\nn~,\\
 &+&\al'\f{\hat{v}_{\bf 78}F_T}{2\sqrt{15}M_*^2} \sum\limits_{i,j=1}^3\[\f{}{} 8y_{ij}^{\pr s} \tl{Q}_L^i(\tl{U}_L^c)^j h_u+(6y_{ij}^a+8y_{ij}^{\pr s})\tl{L}_L^i(\tl{N}_L^c)^j h_u\f{}{}\].
  \eeqa

After ${\bf 650}$ dimensional Higgs acquire VEVs $\langle{\bf 650}\rangle_{\bf (35,1)}$, the new contributions to superpotential
\beqa W  &\supset &\f{\hat{v}_{\bf 650}}{\sqrt{30}M_*} \sum\limits_{i,j=1}^36h_{i,j=1}^a F_{\bf (\bar{5},2)}^i F_{\bf (1,2)}^j H_{\bf(5,1)}\f{}{}~,\nn\\
 &+&\f{\hat{v}_{\bf 650}}{\sqrt{30}M_*} \sum\limits_{i,j=1}^3\(h_{ij}^{\pr s} F_{\bf (10,1)}^i F_{\bf (10,1)}^j H_{\bf (5,1)}\.\nn\\&&~~~~~+2h_{ij}^{\pr s} F_{\bf (10,1)}^i F_{\bf (\bar{5},2)}^j H_{\bf(\bar{5},2)}-4h_{ij}^{\pr s} F_{\bf (\bar{5},2)}^i F_{\bf (1,2)}^j H_{\bf(5,1)}\left.\f{}{}\)~,\nn\\
 &+&\f{\hat{v}_{\bf 650}}{\sqrt{30}M_*} \sum\limits_{i,j=1}^3\(-2h_{ij}^{\pr\pr s} F_{\bf (10,1)}^i F_{\bf (10,1)}^j H_{\bf (5,1)}\.\nn\\&&~~~~~+2h_{ij}^{\pr\pr s} F_{\bf (10,1)}^i F_{\bf (\bar{5},2)}^j H_{\bf(\bar{5},2)}
-4h_{ij}^{\pr\pr s} F_{\bf (\bar{5},2)}^i F_{\bf (1,2)}^j H_{\bf(5,1)}\left.\f{}{}\)~,\nn\\
 &\supset& \f{\hat{v}_{\bf 650}}{\sqrt{30}M_*} \sum\limits_{i,j=1}^3\[\f{}{}(2h_{ij}^{\pr s}-4h_{ij}^{\pr\pr s}) Q_L^i(U_L^c)^j h_u\.\nn\\&&~~~~~~~~~~+(6h_{ij}^a -4h_{ij}^{\pr s}-4h_{ij}^{\pr\pr s})L_L^i(N_L^c)^j h_u\left.\f{}{}\]\nn~,\\
 &+&\f{\hat{v}_{\bf 650}}{\sqrt{30}M_*} \sum\limits_{i,j=1}^3\[\f{}{}(2h_{ij}^{\pr s} +2h_{ij}^{\pr\pr s})\{Q_L^i(D_L^c)^j h_d+L_L^i(E_L^c)^j h_d\}\f{}{}\],
   \eeqa
 while the new contributions to supersymmetry breaking soft trilinear terms
 \beqa
 -{\cal L} &\supset& \al'\f{\hat{v}_{\bf 650}F_T}{\sqrt{30}M_*^2} \sum\limits_{i,j=1}^3\[\f{}{}(2y_{ij}^{\pr s}-4y_{ij}^{\pr\pr s})\tl{Q}_L^i(\tl{U}_L^c)^j h_u\.\nn\\
 &&~~~~~~~+ (6y_{ij}^a -4y_{ij}^{\pr s}-4y_{ij}^{\pr\pr s})\tl{L}_L^i(\tl{N}_L^c)^j h_u \left.\f{}{}\]\nn~,\\
 &+&\al'\f{\hat{v}_{\bf 650}F_T}{\sqrt{30}M_*^2} \sum\limits_{i,j}^3\[\f{}{} (2y_{ij}^{\pr s} +2y_{ij}^{\pr\pr s})\{\tl{Q}_L^i(\tl{D}_L^c)^j h_d
+\tl{L}_L^i(\tl{E}_L^c)^j h_d\}\].
  \eeqa

\section{Scalar and Gaugino Mass Relations}
\label{sec-4}
In order to study the scalar and gaugino mass relations \cite{fei1,Carena:2010gr} that
are invariant under one-loop renormalization group running, we need to
know the renormalization group equations (RGEs) of the supersymmetry
breaking scalar masses and gaugino masses. For simplicity, we only
consider the
one-loop RGE running since the two-loop RGE running
effects are small~\cite{Li:2010mr}.
In particular, for the first two generations, we can
neglect the contributions from the Yukawa coupling terms and trilinear
soft terms, and then the RGEs for the scalar masses
are~\cite{Martin:1993zk}
\beqa 16\pi^2\f{d m^2_{\tl{Q}_{j}}}{d
t}&=&-\f{32}{3}g_3^2M_3^2-6g_2^2M_2^2-\f{2}{15}g_1^2M_1^2+\f{1}{5}g_1^2S~,\\
16\pi^2\f{d m^2_{\tl{U}^c_{j}}}{d
t}&=&-\f{32}{3}g_3^2M_3^2-\f{32}{15}g_1^2M_1^2-\f{4}{5}g_1^2S~,\\
16\pi^2\f{d m^2_{\tl{D}^c_{j}}}{d
t}&=&-\f{32}{3}g_3^2M_3^2-\f{8}{15}g_1^2M_1^2+\f{2}{5}g_1^2S~,\\
16\pi^2\f{d m^2_{\tl{L}_{j}}}{d t}&=&
-6g_2^2M_2^2-\f{6}{5}g_1^2M_1^2-\f{3}{5}g_1^2S~,\\
16\pi^2\f{d m^2_{\tl{E}^c_{j}}}{d
t}&=&-\f{24}{5}g_1^2M_1^2+\f{6}{5}g_1^2S~, \eeqa
where $j=1,~2$, and $t={\rm ln}\mu$ and $\mu$ is the renormalization
scale.
Also, $S$ is given by
\beqa\small
S&=&Tr[Y_{\phi_i}
m^2(\phi_i)]\nn\\&=&m_{H_u}^2-m_{H_d}^2+Tr[M_{\tl{Q}_i}^2-M_{\tl{L}_i}^2
-2M_{\tl{U}^c_i}^2+M_{\tl{D}_i^c}^2+M_{\tl{E}_i^c}^2]~.~\,
\eeqa\normalsize

The one-loop RGEs for gauge couplings $g_i$ and gaugino masses
$M_i$ are
\beqa
\f{d}{dt}g_i~=~\f{1}{16\pi^2}b_ig_i^3~,~~~
\f{d}{dt}M_i~=~\f{1}{8\pi^2}b_ig_i^2M_i~,
\eeqa
where $g_1\equiv \sqrt{5} g_Y/\sqrt{3}$, and $b_1$, $b_2$ and $b_3$
are one-loop beta functions for $U(1)_Y$, $SU(2)_L$, and $SU(3)_C$,
respectively. For the supersymmetric SM, we have
\beqa
b_3=-3~,~b_2=1~,~b_1=\f{33}{5}~.
\eeqa
Therefore, we obtain
\beqa\small
\f{d}{dt} \[\f{MSQj}{Y_{Q_j}}\] &=&\f{d}{dt} \[\f{MSUj}{Y_{U^c_j}}\]
~=~\f{d}{dt} \[\f{MSDj}{Y_{D^c_j}}\] \nonumber \\
&=&\f{d}{dt} \[\f{MSLj}{Y_{L_j}}\]
 ~=~\f{d}{dt} \[\f{MSEj}{Y_{E^c_j}}\]~,
\eeqa\normalsize
where
\beqa
MSQj &=& 4 m^2_{\tl{Q}_{j}} + \f{32}{3b_3} M_3^2 + \f{6}{b_2} M_2^2
+ \f{2}{15b_1} M_1^2 ~,~ \\
MSUj &=& 4 m^2_{\tl{U}^c_{j}} + \f{32}{3b_3} M_3^2
+ \f{32}{15b_1} M_1^2 ~,~ \\
MSDj &=& 4 m^2_{\tl{D}^c_{j}} + \f{32}{3b_3} M_3^2
+ \f{8}{15b_1} M_1^2 ~,~ \\
MSLj &=& 4 m^2_{\tl{L}_{j}} +  \f{6}{b_2} M_2^2
+ \f{6}{5b_1} M_1^2 ~,~ \\
MSEj &=& 4 m^2_{\tl{E}^c_{j}}  + \f{24}{5b_1} M_1^2 ~.~
\eeqa
In addition, we obtain the most general scalar and gaugino mass relations
that are valid from the GUT scale to the electroweak
scale under one-loop RGE running for the first two families
\beqa\small
\gamma_{Q_j} \f{MSQj}{Y_{Q_j}} + \gamma_{U^c_j} \f{MSUj}{Y_{U^c_j}}
+ \gamma_{D^c_j} \f{MSDj}{Y_{D^c_j}} + \gamma_{L_j} \f{MSLj}{Y_{L_j}}
+ \gamma_{E^c_j} \f{MSEj}{Y_{E^c_j}} = C_o,\nn\\
\eeqa\normalsize
where $C_o$ denotes the invariant constant under one-loop
RGE running, and
 $\gamma_{Q_j}$, $\gamma_{U^c_j}$, $\gamma_{D^c_j}$,
$\gamma_{L_j}$, and $\gamma_{E^c_j}$ are real or complex numbers that
satisfy
\beqa
\gamma_{Q_j} + \gamma_{U^c_j} + \gamma_{D^c_j} + \gamma_{L_j}
+ \gamma_{E^c_j} ~=~ 0~.~\,
\eeqa

In short, we can obtain the scalar and gaugino mass relations
that are valid from the GUT scale to the electroweak scale
at one loop. Such relations will be useful to distinguish
between the mSUGRA and GmSUGRA scenarios.

The scalar and gaugino mass relations can be simplified by the
scalar and gaugino mass relations at the GUT scale. Because the
higher-dimensional operators can contribute to gauge kinetic
functions after GUT symmetry breaking,
the SM gauge couplings may not be unified at the
GUT scale. Thus, we will have two contributions
to the gaugino masses at
the GUT scale: the universal gaugino masses as in the mSUGRA,
and the non-universal gaugino masses due to the higher-dimensional
operators.
In particular, for the scenarios studied in
Refs.~\cite{Anderson:1999uia, Chamoun:2001in, Chakrabortty:2008zk,
Martin:2009ad, Bhattacharya:2009wv, Feldman:2009zc, Chamoun:2009nd}
where the universal gaugino masses are assumed to be zero,
{\it i.e.}, $M_i/\alpha_i=a_i M'_{1/2}$,
we obtain the gaugino mass relation at one loop~\cite{Li:2010xr}
\beqa
\f{M_3}{a_3 \alpha_3}=\f{M_2}{a_2 \alpha_2}=\f{M_1}{a_1 \alpha_1} ~.
\eeqa
We can calculate the scalar and gaugino mass relations
in the mSUGRA and GmSUGRA scenarios, and
compare them in different cases.

The RGE running invariant combinations in SU(5), SO(10), Pati-Salam model had been discussed in our previous works \cite{fei1}.
We only discuss here the $SU(3)_C\tm SU(3)_L\tm SU(3)_R$ case from $E_6$ breaking.

We consider the following $E_6$ gauge symmetry breaking chain
\beqa
E_6&\ra& SU(3)_C\tm SU(3)_L\tm SU(3)_R\nn\\&\ra& SU(3)_C\tm SU(2)_L\tm
SU(2)_R\tm U(1)_{B-L}\nn\\&\ra& SU(3)_C\tm SU(2)_L\tm U(1)_Y~. \eeqa
Other symmetry breaking chains can be discussed similarly.

Let us explain our convention. We denote the gauge couplings for
the $SU(2)_L$, $SU(2)_R$, $SU(3)_L$, $SU(3)_R$, $U(1)_{B-L}$ and $SU(3)_C$
gauge symmetries as $g_{2L}$,
$g_{2R}$, $g_{3L}$,
$g_{3R}$, $\tl{g}_{B-L}$ (or traditional
$g_{B-L}$), and $g_3$, respectively.
We denote the gaugino masses for the
$SU(2)_L$, $SU(2)_R$, $SU(3)_L$, $SU(3)_R$, $U(1)_{B-L}$, and $SU(3)_C$ gauge symmetries as
$M_{2L}$, $M_{2R}$, $M_{3L}$, $M_{3R}$, $M_{B-L}$, and $M_3$, respectively.
We denote the one-loop beta functions for the
$SU(2)_L$, $SU(2)_R$, $SU(3)_L$, $SU(3)_R$, $U(1)_{B-L}$, and $SU(3)_C$ gauge symmetries as
$b_{2L}$, $b_{2R}$, $b_{3L}$, $b_{3R}$, $\tl{b}_{B-L}$ and $b_4$, respectively.
In addition, we denote the universal supersymmetry breaking
scale as $M_S$, the $SU(2)_R\times U(1)_{B-L}$ gauge
symmetry breaking scale as $M_{LR}$, and the $SU(3)_C\tm SU(3)_L\tm SU(3)_R$ gauge
symmetry breaking scale as $M_{33}$. Also, we denote
the $U(1)_{B-L}$ charge for the particle $\phi_i$ as $Y^{B-L}_{\phi_i}$.

Neglecting the Yukawa coupling terms and trilinear soft terms,
we obtain the RGEs for the scalar masses of the first two
generations in the gauge group $SU(3)_C\tm SU(3)_L\tm SU(3)_R$ 
\beqa 16\pi^2\f{d m^2_{\tl{X}_L}}{dt}
&=&4\pi^2\f{d}{dt}\[-\f{32}{3b_3}M_3^2-\f{32}{3b_{3L}}M_{3L}^2\]~,\\
16\pi^2\f{d
m^2_{\tl{X}_L^c}}{dt}&=&4\pi^2\f{d}{dt}\[-\f{32}{3b_3}M_3^2-\f{32}{3b_{3R}}M_{3R}^2\]~,\\
16\pi^2\f{d m^2_{\tl{N}}}{dt}&=&4\pi^2\f{d}{dt}\[-\f{32}{3b_{3L}}M_{3L}^2-\f{32}{3b_{3R}}M_{2R}^2\]~,
\eeqa
which gives \beqa
\f{d}{dt}\[m^2_{\tl{X}_L}+\f{8}{3b_3}M_3^2+\f{8}{3b_{3L}}M_{3L}^2\]&=&0~,\\
\f{d}{dt}\[m^2_{\tl{X}_L^{c}}+\f{8}{3b_3}M_3^2+\f{8}{3b_{3R}}M_{3R}^2\]&=&0~\\
\f{d}{dt}\[m^2_{\tl{N}}+\f{8}{3b_{3L}}M_{3L}^2+\f{8}{3b_{3R}}M_{3R}^2\]&=&0.
\eeqa
The RGEs of the scalar masses for the
first two generations in the left right model 
$SU(3)_C\tm SU(2)_L\tm SU(2)_R\tm U(1)_{B-L}$ are
\beqa\small 16\pi^2\f{d m^2_{\tl{Q}_{j}}}{d
t}&=&-\f{32}{3}g_3^2M_3^2-6g_{2L}^2M_{2L}^2-\f{1}{3}\tl{g}_{B-L}^2M_{B-L}^2+\f{1}{2}\tl{g}_{B-L}^2S^\pr~,\nn\\
16\pi^2\f{d m^2_{\tl{U}_j^c,\tl{D}_{j}^c}}{d
t}&=&-\f{32}{3}g_3^2M_3^2-6g_{2R}^2M_{2R}^2-\f{1}{3}\tl{g}_{B-L}^2M_{B-L}^2-\f{1}{2}\tl{g}_{B-L}^2S^\pr~,\nn\\
16\pi^2\f{d m^2_{\tl{L}_{j}}}{d t}&=&
-6g_{2L}^2M_{2L}^2-3\tl{g}_{B-L}M_{B-L}^2-\f{3}{2}\tl{g}_{B-L}^2S^\pr~,\nn\\
16\pi^2\f{d m^2_{\tl{E}^c_{j}}}{d
t}&=&-6g_{2R}^2M_{2R}^2-3\tl{g}_{B-L}M_{B-L}^2+\f{3}{2}\tl{g}_{B-L}^2S^\pr~,\eeqa\normalsize
where
\beqa
S^\pr=Tr[Y_{\phi_i}^{B-L}m^2(\phi_i)]~.
\eeqa
We consider the following linear combination of the squared scalar masses
\beqa
&&16\pi^2\f{d}{dt}\(m_{\tl{U}_j^c}^2+m_{\tl{E}_j^c}^2-2m_{\tl{Q}_j}^2\)
\nonumber \\
=&&4\pi^2\f{d}{dt}\[\f{32}{3b_3}M_3^2+\f{12}{b_{2L}}M_{2L}^2-\f{20}{3{b}_{1}}
M_{1}^2\]~~~{\rm for }~~M_{S}<\mu<M_{LR} \nonumber \\
=&&4\pi^2\f{d}{dt}\[\f{32}{3b_3}M_3^2
+\f{12}{b_{2L}}M_{2L}^2-\f{12}{b_{2R}}M_{2R}^2-\f{8}{3\tl{b}_{B-L}}
M_{B-L}^2\] \nonumber \\ &&
~~~{\rm for }~~M_{LR}<\mu<M_{33} \nonumber \\
=&&4\pi^2\f{d}{dt}\[-\f{64}{3b_{3R}}M_{3R}^2+\f{32}{3b_{3L}}M_{3L}^2+\f{32}{3b_{3}}M_{3}^2\]
~~~{\rm for }~~M_{33}<\mu<M_{U}~.~\,\nn\\
\eeqa
From the RGE invariant combinations, we obtain the one-loop exact scalar and gaugino mass relations 
from the GUT scale to the electroweak scale 
\beqa\tiny
&&4\(m_{\tl{U}_j^c}^2+m_{\tl{E}_j^c}^2-2m_{\tl{Q}_j}^2\)
-\f{32M_3^2}{3b_3}-\f{12M_{2L}^2}{b_{2L}}+\f{20M_1^2}{3b_1}=C_o^1~,\nn\\
&&4\(m_{\tl{U}_j^c}^2+m_{\tl{E}_j^c}^2-2m_{\tl{Q}_j}^2\)-\f{32M_3^2}{3b_3}
-\f{12M_{2L}^2}{b_{2L}}+\f{12M_{2R}^2}{b_{2R}}
+\f{8M_{B-L}^2}{3\tl{b}_{B-L}} =C_o^2~,\nn\\
&&4\(m_{\tl{U}_j^c}^2+m_{\tl{E}_j^c}^2-2m_{\tl{Q}_j}^2\)-\f{32M_{3}^2}{3b_{3}}+\f{64M_{3R}^2}{3b_{3R}}
-\f{32M_{3L}^2}{3b_{3L}}=C_o^3~.
\eeqa\normalsize
The differences between the constants $C_o^1$ and $C_o^2$ and
between the constants $C_o^2$ and $C_o^3$ are the
threshold contributions from the extra particles due to gauge
symmetry breaking. Thus, the three constants can be determined by
matching the threshold contributions at the symmetry breaking scales.
The difference between $C_o^2$ and $C_o^3$ is
\beqa\tiny
C_o^2-C_o^3
&=&\(\f{12}{b_{2R}}-\f{64}{3b_{3R}}\)M_{3R}^2+ \(\f{32}{3b_{3L}}-\f{12}{b_{2L}}\)M_{3L}^2+\f{8}{3\tl{b}_{B-L}}M_{B-L}^2~,\nn\\
\eeqa\normalsize
while the difference between $C_o^1$ and $C_o^2$ is
\beqa
C_o^1-C_o^2
&=&-\f{12}{b_{2R}}M_{2R}^2-\f{8}{3\tl{b}_{B-L}}M_{B-L}^2+\f{20}{3b_{1}} M_{1}^2~.
\eeqa

  At the $SU(3)_C\tm SU(3)_L\times
SU(3)_{R}$ unification scale $M_{33}$, we have
\beqa
\f{1}{g_{B-L}^2}=\f{1}{g_{3L}^2}+\f{1}{g_{3R}^2}~.
\eeqa

For mSUGRA with universal gaugino and scalar masses, we have
\beqa
\f{M_3}{g_3^2}=\f{M_{2L}}{g_{2L}^2}=\f{M_{2R}}{g_{2R}^2}=\f{M_{3L}}{g_{3L}^2}=\f{M_{3R}}{g_{3R}^2}~.
\eeqa
Thus, we can get the scalar and gaugino mass relations in supersymmetric
Standard Model
\beqa
&&4\(m_{\tl{U}_j^c}^2+m_{\tl{E}_j^c}^2-2m_{\tl{Q}_j}^2\)
-\f{32}{3b_3}M_3^2-\f{12}{b_{2L}}M_2^2+\f{20}{3b_1}M_1^2\nn\\
&=& \(2\f{8}{3\tl{b}_{B-L}}-\f{32}{3b_{3L}}\)\f{M_3^2(\mu)}{g_3^4({\mu})}g_3^4(M_{33})
+\f{20}{3b_1}\f{M_1^2(\mu)}{g_1^4({\mu})}g_1^4(M_{LR})\nn\\
&&~~-\(\f{12}{b_{2R}}g_{2R}^4(M_{LR})+
\f{8}{3\tl{b}_{B-L}}g_{B-L}^4(M_{LR})\)\f{M_3^2(\mu)}{g_3^4({\mu})}~.\eeqa
Here we use the fact that $b_{3}=b_{3L}=b_{3R}$ for $(M_{E_6}>\mu>M_{33})$ as well as $b_{2L}=b_{2R}$ for $(M_{33}>\mu>M_{LR})$.
If we know the low energy sparticle spectrum at the LHC and ILC
and $g_1^2(M_{LR})$ from the RGE running, we can get the coefficients
\beqa
c&=&\(\f{16}{3\tl{b}_{B-L}}-\f{32}{b_3}\)g_3^4(M_{PS})-\(\f{12}{b_{2R}}g_{2R}^4(M_{LR})+
\f{8}{3\tl{b}_{B-L}}g_{B-L}^4(M_{LR})\),\nn\\ \eeqa by fitting the
experimental data.

For GmSUGRA with non-universal gaugino and scalar masses, we consider the
Higgs field in the ${\bf 650}$ representation whose singlet component
$({\bf 1,1,1})$ acquires VEVs. To give mass to the gluino, we require
that the universal gaugino mass be non-zero.
From Eq.~(\ref{33sm-6501}), we obtain
\beqa
m_{\tl{E}_j^c}^2+m_{\tl{U}_j^c}^2-2m_{\tl{Q}_j}^2=-\f{\sqrt{2}}{2}({\beta'}^{\bf
650}v_{\bf 650})
\f{|F_S|^2}{M_*^3}~.
\eeqa
Thus, the constant combination in the supersymmetric Standard Model is
\beqa
&&4\(m_{\tl{U}_j^c}^2+m_{\tl{E}_j^c}^2-2m_{\tl{Q}_j}^2\)-\f{32}{3b_3}M_3^2-\f{12}{b_{2L}}M_{2L}^2
+\f{20}{3b_1}M_1^2\nn\\
&=&-\f{\sqrt{2}}{2}({\beta'}^{\bf
650}v_{\bf 650})
\f{|F_S|^2}{M_*^3}+
\f{20}{3b_1}\f{M_1^2(\mu)}{g_1^4({\mu})}g_1^4(M_{LR})\nn\\&-&\(\f{12}{b_{2R}}g_{2R}^4(M_{LR})
\f{M_{2R}^2(\mu)}{g_{2R}^4({\mu})}+
\f{8}{3\tl{b}_{B-L}}g_{B-L}^4(M_{LR})\f{M_3^2(\mu)}{g_3^4({\mu})}\)\nn\\
&+&\(\f{16}{3\tl{b}_{B-L}}-\f{32}{3b_{3L}}\)\f{M_3^2(\mu)}{g_3^4({\mu})}g_3^4(M_{33})~.
\eeqa
Therefore, the scalar and gaugino mass relations
in mSUGRA are different from those in GmSUGRA.
Similar discussions can be used for other $E_6$ gauge symmetry breaking chains and we will not present here.

\section{Conclusions}
\label{sec-5}
In the GmSUGRA scenario with the
higher-dimensional operators containing the GUT Higgs fields,
we systematically studied the supersymmetry breaking scalar masses,
SM fermion Yukawa coupling terms,
and trilinear soft terms in the $E_6$ model where
the gauge symmetry is broken down to the
$SO(10)\tm U(1)$ gauge symmetry,
$SU(3)_C\times SU(3)_L \times SU(3)_R$
gauge symmetry, $SU(6)\times SU(2)_a (a={\rm L,R,X})$
gauge symmetry, flipped $SU(5)$ gauge symmetry. In addition, we considered the scalar and
gaugino mass relations, which can be preserved from
the GUT scale to the electroweak scale
under one-loop RGE running, in the $SU(3)_C\times SU(3)_L \times SU(3)_R$
model arising from the $E_6$ model. With such relations, we may distinguish the
mSUGRA and GmSUGRA scenarios if we can measure the supersymmetric
particle spectrum at the LHC and ILC.
Thus, it provides us with another important window of opportunity at the Planck scale.




\section*{Acknowledgments}
We are very grateful to the referee for very useful suggestions and helps.
This research was supported by the Australian Research
Council under project DP0877916 and by the Natural Science Foundation
of China under project "The Low Energy Effects of Non-Renormalizable Terms in GUT Models".



\bibliographystyle{model1a-num-names}

\end{document}